\documentclass{aa}

\usepackage[T1]{fontenc}
\usepackage{txfonts}
\usepackage{amsmath}
\usepackage{amsfonts}
\usepackage{xcolor}
\usepackage{url}
\usepackage{bm}
\usepackage{graphicx}
\usepackage{amssymb}
\usepackage{soul}
\usepackage{booktabs}
\usepackage{array}
\usepackage[utf8]{inputenc}
\inputencoding{latin1}
\inputencoding{utf8}
\usepackage{appendix}
\usepackage{placeins}
\usepackage{multicol}

\DeclareMathOperator\erf{erf}

\def\lsim{ \lower .75ex\hbox{$\sim$} \llap{\raise .27ex \hbox{$<$}} }
\def\gsim{ \lower .75ex \hbox{$\sim$} \llap{\raise .27ex \hbox{$>$}} }

\newcommand{\fig}[1]{Fig.~\ref{fig:#1}}
\newcommand{\eq}[1]{Eq.~(\ref{eq:#1})}

\newcommand{\unit}[1]{\nobreak{\mathrm{\;#1}}}

\newcommand{\bi}{\begin{itemize}}
\newcommand{\ei}{\end{itemize}}

\usepackage[colorlinks=true,linkcolor=blue,citecolor=blue]{hyperref}

\title{The polarization of the synchrotron radiation from a recollimated jet: Application to high-energy BL Lacs} 

\author{
A. Sciaccaluga\inst{1,2}
\and 
A. Costa\inst{3}
\and
F. Tavecchio\inst{1}
\and
G. Bodo\inst{3}
\and
P. Coppi\inst{4}
\and
S. Boula\inst{1}
}

\institute{
INAF -- Osservatorio Astronomico di Brera, Via E. Bianchi 46, I-23807 Merate, Italy\\
\email{alberto.sciaccaluga@inaf.it}
\and
Dipartimento di Fisica, Università degli Studi di Genova, Via Dodecaneso 33, I-16146 Genova, Italy
\and
INAF -- Osservatorio Astrofisico di Torino, Strada Osservatorio 20, 10025 Pino Torinese, Italy
\and
Department of Astronomy, Yale University, PO Box 208101, New Haven, CT 06520-8101, USA
}
\date{}

\titlerunning{Polarization of synchrotron radiation from recollimating jet}

\begin{document}


\abstract{Multifrequency polarimetry, recently extended to the X-ray band thanks to the Imaging X-ray Polarimetry Explorer (IXPE) satellite, is an essential tool for understanding blazar jets. High-frequency-peaked BL Lacs (HBLs) and extreme high-frequency-peaked BL Lacs (EHBLs) are especially interesting because the polarimetric properties of their synchrotron emission, extending up to the X-ray band, can be fully tracked by sensitive polarimetric measurements. We investigated the polarization properties of the synchrotron emission of these sources, starting directly from relativistic magnetohydrodynamic simulations of recollimated relativistic jets. To bridge the gap between fluid and kinetic scales, we elaborated a post-processing code based on the Lagrangian macroparticle approach, which models the spectral evolution and emission of nonthermal particles within the jet given the local fluid conditions. When comparing our results with early particle-in-cell (PIC) simulations, we find that shocks formed through jet recollimation are primarily superluminal, limiting particle acceleration in a laminar flow. However, recent PIC simulations suggest that acceleration can occur in the presence of small-scale turbulence or inhomogeneities even in a superluminal configuration. In this case, we reproduce the observed polarization chromaticity (i.e., the polarization degree increases with frequency), along with a stable polarization angle between the X-ray and optical bands. This study sheds light on the role of recollimation shocks in blazar jets and supports the energy-stratified shock model as a plausible explanation for IXPE observations.}


\maketitle

\section{Introduction}

Multifrequency polarimetry is emerging as a powerful probe of blazar jets, especially with the advent of the Imaging X-ray Polarimetry Explorer (IXPE) satellite \citep{weisskopf+16}. IXPE has opened a new window for polarimetry in the X-ray band, providing unprecedented insights into the behavior of these extreme astrophysical objects. Blazars at the low-power end of the blazar sequence, i.e., the most efficient particle accelerators, are excellent targets for IXPE because their emission peaks in the X-ray band \citep{tavecchio21}. This is particularly true for High-frequency-peaked BL Lacs (HBLs) and Extreme High-frequency-peaked BL Lacs (EHBLs), where both optical and X-ray emissions can be attributed to synchrotron radiation from a population of nonthermal electrons. 

Several multiwavelength polarimetric campaigns have been organized to measure the polarization properties of HBLs and EHBLs across the electromagnetic spectrum (e.g., \citealt{liodakis+22}, \citealt{digesu+22}, and \citealt{ehlert+23}). Observations of blazars in a quiescent state, where the polarization degree and the electric vector position angle (EVPA) do not show significant temporal variability, have revealed the following key characteristics. First, the polarization degree increases with frequency, a phenomenon known as chromaticity. Specifically, the observed X-ray polarization degree is typically $\Pi_x \sim 10-20\%$, which is significantly higher than the optical and radio polarization degree, with $\Pi_x/\Pi_o \sim 2-6$, where $\Pi_o$ is the optical polarization degree. Such strong chromaticity suggests that the mechanisms influencing polarization are energy-dependent. Second, the EVPA is weakly chromatic, with the X-ray and optical EVPAs being approximately aligned ($\psi_x \sim \psi_o$), indicating a stable magnetic field orientation in the emitting region. Moreover, the EVPA tends to align with the projection of the jet axis on the plane of the sky. However, this alignment is not consistent across all sources and comes with certain caveats (more details in the discussion). Finally, IXPE also observed an episode of temporal variability, specifically a rotation of the X-ray polarization angle observed in the HBL Mrk421 in June 2022 by more than $360^\circ$ in $5$ days \citep{digesu+23}. 

Several models have been proposed to explain the recent results on the multiwavelength polarization of HBLs and EHBLs. For instance, some models are based on the global structure of the jet magnetic field (e.g., \citealt{bolis+24}). Currently, the most accepted scenario is the energy-stratified shock model \citep{angelakis+16, tavecchio+18}. Supposing that nonthermal electrons are accelerated by a shock, those emitting X-ray photons cool down in a small region close to the shock front. In this region, the magnetic field is ordered by shock compression and there is a strong self-generated component, implying a large degree of polarization. On the other hand, the particles emitting optical and radio photons cool down in a larger region and are advected by the flow farther downstream. In this case, electrons are subject to several radiation zones with different magnetic field orientations, implying a low polarization degree. A likely origin for the shock responsible for blazar emission is the recollimation of the jet by external gas \citep{tagliaferri+08,marscher14,zech21,tavecchio22,costa23} and this is the scenario that we adopt in the following.

Our goal is to give a more quantitative description of the energy-stratified model in the framework of a recollimating jet scenario, starting directly from relativistic magnetohydrodynamic (RMHD) simulations. From fluid simulations, we obtained the large-scale structure of the magnetic and velocity field, necessary to estimate the nonthermal particle emission and its beaming. However, fluid simulations cannot resolve kinetic scales and provide no information about the injection and acceleration of nonthermal particles, which requires tracking particles on scales of the electron gyroradius, as done in particle-in-cell (PIC) simulations.

One possible solution to bridge the gap between fluid and kinetic scales is the Lagrangian macroparticle approach. A macroparticle is an ensemble of nonthermal particles decoupled from the fluid. The macroparticles move along the streamlines, and their spectra are updated using the local fluid conditions. The particles sample the fluid, and their integrated emission is used to estimate the radiative output of the jet \citep{mimica+09, mimica&aloy12}. Relativistic hydrodynamic simulations have frequently been considered because of their lower computational cost. In this case, the magnetic field is obtained assuming an equipartition between the magnetic and internal energy density. Moreover, it was standard to consider a fixed power law index for the spectra of the macroparticle injected at the shock, neglecting the dependence of the acceleration process on the shock features. 

These two limitations have been surpassed by the approach outlined in \cite{vaidya+18}. This hybrid framework estimates the emission directly from RMHD simulations and implements a sub-grid model for shock acceleration. The magnetic field is used to calculate the synchrotron losses and emission of the macroparticles, accounting for beaming and the relativistic aberration of synchrotron polarization. Moreover, when macroparticles cross a shock, their spectra are updated using the local fluid conditions, specifically the magnetic orientation and the shock compression ratio. 

In this paper, we implemented a new post-processing code, based on the work of \cite{vaidya+18} and \cite{mukherjee+21}, which can calculate the radiative output of stationary RMHD simulations. Comparing our results with early PIC simulations, we find that shocks formed by jet recollimation are predominantly superluminal, which restricts particle acceleration in a laminar flow (\citealt{crumley+19} for the transrelativistic case, \citealt{sironi&spitkovsky09, sironi&spitkovsky11} for the relativistic case). However, recent PIC simulations indicate that acceleration can still occur in a superluminal configuration if small-scale turbulence or inhomogeneities are present \citep{bresci+23, demidem+23}. With particle acceleration turned on, we focused on investigating the energy-stratified shock scenario, specifically evaluating whether it can replicate the polarimetric features for HBLs and EHBLs detected by IXPE and the simultaneous multiwavelength campaigns. In Sect. \ref{sec:RMHD setup} we describe the RMHD simulations and in Sect. \ref{sec:post processing} the post-processing code. The application is reported in Sect. \ref{sec:results} and in Sect. \ref{sec:discussion} we discuss the results. 

\section{RMHD setup}
\label{sec:RMHD setup}

\subsection{Context}
We simulated the recollimation of a relativistic jet at parsec-scale distance from the supermassive black hole, where the high-energy emission region is probably located \citep{tavecchio+10}. Aside from occasional flares, the observed emission remains stable throughout day-long observation periods. This level of stability would be difficult to explain if the emission were generated by a shock traveling through the jet. Assuming that recollimation and reflection shocks are responsible for particle acceleration is not a standard choice since in blazar modeling it is often postulated that the emission region is downstream of a shock of normal incidence traveling down the jet. Possible physical realizations of this setup could be represented by a dense blob of plasma overtaking the jet plasma or by internal shocks developing between portions of the flow characterized by different speeds \citep{rees78, spada+01}. However, a difficulty of such a setup (already pointed out by e.g., \citealt{tagliaferri+08}) is that in an observed time interval $\Delta t$, the shock moves by a distance $\Delta z\sim c \Delta t \Gamma_{\rm sh, obs}^2$, where $\Gamma_{\rm sh, obs}$ is the Lorentz factor of shock in the observer frame. Assuming typical values, namely $\Delta t = 1 \unit{day}$ and $\Gamma_{\rm sh, obs} = 30$, the distance amounts to $\sim 4 \unit{pc}$. During the propagation (and the consequent expansion) the physical parameters of the plasma in the jet (density, magnetic field) would drastically change, greatly modifying the emission properties \citep{blandford&konigl79}. Such a conclusion is in patent disagreement with the stability of the emission during the considered observation periods (e.g., \citealt{liodakis+22}). 
We remark that the large Lorentz factor of the shock in the observer frame is a direct consequence of the fact that to properly model blazar emission we need to Doppler boost the emission from the downstream plasma, while for a relativistic shock of normal incidence, the velocity of the downstream flow in the shock frame is necessarily sub-relativistic, hence the shock frame must move relativistically in the observer frame. This problem can be overcome in the case of an oblique shock, since in that case only the normal component of the downstream velocity must be sub-relativistic with respect to the shock. In this case, the shocked plasma can maintain a substantial bulk Lorentz factor with respect to the observer even if the shock is stationary in that frame. The low temporal variability can be related to some permanent and fixed (in the observer frame) structure in the jet, an oblique standing recollimation region, shaped by the pressure contrast between the jet and the external medium, e.g., \cite{komissarov&falle97} and \cite{bodo&tavecchio18}. Moreover, recollimation shocks could also be a natural outcome of MHD jets even without external confinement, e.g., \cite{jannaud+23}. 

We started from the axisymmetric hydrodynamic setup described in \cite{costa23}, but we included the magnetic field. We considered small magnetization values, for which the jet is expected to be unstable (\citealt{matsumoto21}, Boula et al. in prep.) against recollimation instabilities. However, the first series of recollimation-reflection shocks is not disrupted \citep{costa23}.

\subsection{Numerical framework}

Following \cite{costa23}, we looked for an axisymmetric steady-state solution by performing relativistic magnetohydrodynamic simulations until stationarity was reached. We used the ``RMHD'' module of the state-of-the-art PLUTO code for astrophysics \citep{mignone07,mignone12}, which solves the equations of ideal RMHD
\begin{equation}
    \frac{\partial }{\partial t} \left( \begin{matrix}
        d\\
        \mathbf{m}\\
        e_t\\
        \mathbf{B}
    \end{matrix} \right) + \boldsymbol{\nabla}\cdot \left( \begin{matrix}
        d\mathbf{v}\\
        \omega_t\Gamma^2\mathbf{v}\mathbf{v}-\mathbf{b}\mathbf{b}+p_{t}\mathbf{I}\\
        \mathbf{m}\\
        \mathbf{v}\mathbf{B}-\mathbf{B}\mathbf{v}
    \end{matrix}\right) = \left( \begin{matrix}
        0\\
        \mathbf{f}_g\\
        \mathbf{f}_g\cdot \mathbf{v}\\
        \mathbf{0}
    \end{matrix} \right),
    \label{eq:pluto_rhd}
\end{equation}
where the speed of light $c$ is set to $1$, and a factor $\sqrt{4\pi}$ is reabsorbed in the definition of $\mathbf{B}$.
The set of primitive variables consists of the rest frame density, the thermal pressure, the three-velocity in the lab frame, and the magnetic field three-vector in the lab frame  $(\rho,p,\mathbf{v},\mathbf{B})$. $\Gamma$ is the Lorentz factor, $\mathbf{f}_g$ is the force density in the lab-frame, and $\mathbf{I}$ is the unit $3 \times 3$ tensor. The conserved quantities are $\mathbf{B}$, and the laboratory frame mass, momentum, and energy densities, respectively defined as
\begin{equation}
    \begin{matrix}
        d=\rho\Gamma\\
        \mathbf{m}=w_t\Gamma^2\mathbf{v}-b^0\mathbf{b}\\
        e_t = w_t\Gamma^2-b^0b^0-p_t
    \end{matrix},\quad\mbox{with}\quad
    \begin{matrix}
       & b^0=\Gamma\mathbf{v}\cdot\mathbf{B}\\ &\mathbf{b}=\mathbf{B}/\Gamma+\Gamma\left(\mathbf{v}\cdot\mathbf{B}\right)\mathbf{v}\\
        &w_t=\rho h + \mathbf{B}^2/\Gamma^2+\left(\mathbf{v}\cdot\mathbf{B}\right)^2\\
        &p_t = p + \frac{\mathbf{B}^2/\Gamma^2+\left(\mathbf{v}\cdot\mathbf{B}\right)^2}{2}
    \end{matrix}.
\end{equation}
The system of RMHD equations is closed by the Taub-Matthews equation of state, defining the rest frame relativistic specific enthalpy as \citep{mignoneTM}
\begin{equation}
    h = \frac{5}{2}\Theta+\sqrt{\frac{9}{4}\Theta^2+1},
\end{equation}
with $\Theta=p/(\rho c^2)$, approximating the Synge EoS of a single-specie relativistic perfect fluid \citep{synge}. 
 
The numerical methods included a linear reconstruction scheme \citep{mignone07}, second-order Runge-Kutta time integration \citep{butcher_rk}, and the HLLD Riemann solver \citep{pluto_hlld}. We adopted a constrained transport approach, which ensures that the divergence of the magnetic field is null throughout the simulation \citep{pluto_ct}.

The simulation was carried out in 2D cylindrical coordinates $(R,z)\in[0,6]\times[1,30]$ in units of $z_0$, defined below, with $1400\times2200$ points. Boundary conditions were set to outflow at all boundaries except at the axis, where they were reflective, and at the base, at $z=z_0$. There, analytical profiles were defined constant to continuously inject the jet at radii $R<R_{j}=\theta_jz_0$, and fixed the environment profile at larger radii too. The steady state obtained from the final output, at $t=3000 t_0,$ where $t_0= z_0/c$, was finally interpolated to a 3D Cartesian domain $(x,y,z)\in[-0.5,0.5]\times[-0.5,0.5]\times[1,10.5]$ with $500\times500\times750$ points uniformly spaced. 
 
\subsection{Parameters and magnetic field}

We simulated a recollimating jet, following the setup described in \cite{costa23}. The hydrodynamic setup was the same: an expanding relativistic conical jet of width $\theta_j$ was initialized in the computational domain at a distance $z_0$ from the cone vertex, surrounded by a confining environment. In these simulations, we further included the magnetic field. 

The magnetic field is helical, defined in cylindrical coordinates in the laboratory frame with components
\begin{align}
    &B_R= \frac{\alpha B_0}{r^2}  \text{e}^{-\mathcal{X}^2} \frac{R}{r},\\
    &B_z = \frac{\alpha B_0}{r^2} \text{e}^{-\mathcal{X}^2} \frac{z}{r},\\
    &B_\phi =  \Gamma \frac{B_0}{r}  \sqrt{\text{e}^{-2\mathcal{X}^2} -\frac{\psi_\chi}{2 \sin^2\theta}\left[\psi_\chi-\psi_\chi \text{e}^{-2\mathcal{X}^2}+\sqrt{2\pi}\,\erf\left(\sqrt{2}\mathcal{X}\right)\right]},
\end{align}
where $r=\sqrt{R^2+z^2}$, and $\mathcal{X} = \left(\cos\theta-1\right)/\psi_\chi$. $\Gamma$ and $\theta=\cos^{-1}(z/r)$ are functions of the position, $\alpha$ and $B_0$ are parameters related to the pitch and to the amplitude of the field in the rest frame; finally, $\psi_\chi$ is the scale length for the decay of the poloidal field, which is assumed to have a Gaussian profile. In our simulations, we set $\psi_\chi=10^{-2}\theta_j$, to make the field decay to $0$ outside the jet.

Inside the computational domain, the poloidal field $B_{\rm pol}=\sqrt{B^2_R+B_z^2}$ peaks in $(0,z_0)$ to $\alpha B_0$, while the toroidal field $B_{\rm tor}=B_\phi$ peaks approximately in $(0.5\theta_j,z_0)$ to $\max(B_{\rm tor}) = 0.7\Gamma B_0$.

These profiles ensure that the magnetic field is in an equilibrium configuration at the lower boundary when $\alpha=1$. In case $\alpha<1$, the jet thermal pressure is defined to compensate for the lower poloidal field,
\begin{equation}
    p_j = p_{j,HD} +(1-\alpha^2) \frac{B^2_0}{2} \frac{\text{e}^{-2\mathcal{X}^2}}{ r^{2\gamma_\text{ad}}},
\end{equation}
where $p_{j,HD}$ is the hydrodynamic adiabatic pressure profile used in \cite{costa23} and $\gamma_\text{ad}=5/3$ is the adiabatic index.
More details on the magnetic field prescription can be found in the appendix \ref{sec:B_details}.

To simulate the properties of HBLs, we constrained the emitting region to be under/around the parsec scale, where the bulk Lorentz factor is expected to be on the order of $10$, the proper frame magnetic field (or better, the component responsible for the emission) around $10^{-2} \unit{G}$, and the power around $10^{43}$ erg/s (e.g., \citealt{tavecchio10}). The jet steady state is determined by the environment, and by its own properties, parameterized by a few quantities that we defined at the injection boundary, which we set at $z_0=0.1$ pc. In general, we use the suffixes ``j'' and ``\text{ext}'' respectively to refer to jet and environment parameters.
We defined $\Gamma_j=10$, the jet's opening angle $\theta_j=0.1$, the jet-enviornment density contrast $\nu= \rho_j(z_0)/\rho_\text{ext}(z_0)=7.6\times10^{-6}$, and we considered a cold jet, with $h_j(z_0)\simeq 1$ (the jet-environment pressure ratio is $p_r=p_j(z_0)/p_\text{ext}(z_0) =10^{-3}$). 

The confining external medium is at rest in the laboratory frame, it is isothermal, with $T = 1.5\times 10^{7}$ K, and it is stratified, following a power law pressure profile, $p_\text{ext}(z) = p_\text{ext}(z_0) \left(z/z_0\right)^{-0.5}$, where $p_\text{ext}(z_0)=3\times 10^{-6}\rho_\text{ext}(z_0)c^2$ and $\rho_{\text{ext}}(z_0)=10^5\,m_p\text{cm}^{-3}$, was chosen to tune the jet power to about $10^{43}$ erg/s. The environment is kept in hydrodynamic equilibrium with a force density $\mathbf{f}_g = \boldsymbol{\nabla}p_\text{ext}$.

For the magnetic field, we set the rest frame jet magnetization at the nozzle, defined as
\begin{equation}
    \sigma_j = \frac{B_0^2}{\rho_{j}(z_0)c^2} = 10^{-2},
    \label{eq:sigma}
\end{equation}
so the proper-frame amplitude of the magnetic field at jet injection is 
\begin{equation}
    B_0 = 1.2 \times10^{-2}\left(\frac{10^{-2}}{\sigma}\right)\left(\frac{\rho_\text{ext}(z_0)}{10^5 m_p\text{cm}^{-3}}\right) \text{G}.
\end{equation}
We studied three cases at different pitches: Case A assumes a force-free field, with $\alpha=1$, Case B considers a toroidal-dominated field, with $\alpha=0.5$, and Case C, where $\alpha = 0$, is for a purely toroidal field.

We finally observed that, since $\sigma_j\ll1$ in all cases we considered, the Poynting flux contribution to the jet power $L_j$ was negligible with respect to the hydrodynamic jet power, $L_{j,HD}$, and we found
\begin{align}
    L_{j} & \simeq L_{j,HD} =  (\pi z_0^2\theta_j^2)\rho_j h_j c^2 \Gamma_j^2 v_j  \\
    \simeq & 10^{43}\left(\frac{\theta_j}{0.1}\right)^2\left(\frac{\nu}{10^{-5}}\right) 
  \left(\frac{\rho_\text{ext}(z_0)}{10^5 m_p\text{cm}^{-3}}\right)\left(\frac{h_j}{1}\right)\left(\frac{\Gamma_j}{10}\right)^2\,\mbox{erg s$^{-1}$}.
\end{align}

\section{Post-processing calculation of particle evolution and emission}
\label{sec:post processing}

As mentioned above, the emission properties of the recollimation shock structure were evaluated in post-processing, tracking particle evolution and emission by means of a Lagrangian macroparticles approach.

A Lagrangian macroparticle is a cloud of actual particles (electrons, protons, etc.) represented by a Dirac delta in physical space and characterized by a nontrivial distribution in phase space. The Lagrangian macroparticles are advected in the Eulerian grid according to
\begin{equation}
    \frac{d\vec{x}_p}{dt} = \vec{v}_p(\vec{x}_p),
    \label{eq:position equation}
\end{equation}
where $\vec{v}_p(\vec{x}_p)$ is the fluid velocity interpolated at the macroparticle position $\vec{x}_p$. In our code, the fluid quantities were interpolated at macroparticle positions using linear interpolation. \eq{position equation} was numerically solved by the second-order Runge-Kutta scheme.

While a macroparticle is moving in the fluid, its spectrum is simultaneously updated by using the local fluid conditions. The spectral time evolution is governed by the relativistic transport equation. For the full equation under the assumption of momentum isotropy, refer to \cite{webb89}. For our application, we can neglect spatial diffusion, shear acceleration, second-order Fermi acceleration, and non-inertial energy change. The transport equation reduces to
\begin{equation}
    p^2 \frac{d f}{d \tau} + \frac{\partial}{\partial p} \left[ 
    -\frac{p^3}{3} f \nabla_\mu u^\mu + \langle \dot{p} \rangle_\text{rad} f 
    \right] = -p^2 f \nabla_\mu u^\mu,
    \label{eq:simplified transport equation}
\end{equation}
where $u^\mu$ is the fluid four-velocity, while all other quantities are expressed in the fluid rest frame. We denote the isotropic phase-space distribution function by $f$ and the Lagrangian derivative with respect to proper time by $d/d\tau = u^\mu \nabla_\mu$. The two terms in the square brackets respectively describe the adiabatic and radiative losses ($\langle \dot{p} \rangle_\text{rad}$ is the average momentum loss due to radiative mechanisms). The right-hand side term accounts for changes in macroparticle number density due to fluid expansion or contraction. When considering a single macroparticle, the total number of particles remains constant. However, as the surrounding fluid expands or contracts, its volume changes, leading to a variation in number density. We define $\mathcal{N}(p,\tau) = \int d \Omega p^2 f = 4 \pi p^2 f$, the number of particles per unit of volume and momentum. Since nonthermal particles are ultrarelativistic, their energy can be expressed as $E=pc$, which in turn implies $\mathcal{N}(E,\tau) = \mathcal{N}(p,\tau)/c$, where $\mathcal{N}(E,\tau)$ is the number of particles per unit volume and energy. Integrating \eq{simplified transport equation} over the solid angle and moving to the energy space gives
\begin{equation}
    \frac{d\mathcal{N}(E,\tau)}{d\tau} + \frac{\partial}{\partial E} \left[ \left( -\frac{E}{3} \nabla_\mu u^\mu + \dot{E} \right) \mathcal{N}(E,\tau) \right] = -\mathcal{N}(E,\tau) \nabla_\mu u^\mu,
    \label{eq:simplified transport equation 2}
\end{equation}
where $\dot{E}_r = \langle \dot{p} \rangle_\text{rad}/p^2$. Finally, it is possible to eliminate the term on the right side by introducing $\chi = \mathcal{N}(E,\tau)/n$, where $n$ is the fluid number density. Using particle number conservation, \eq{simplified transport equation 2} can be rewritten as
\begin{equation}
    \frac{d\chi}{d\tau} + \frac{\partial}{\partial E} \left[ \left( -\frac{E}{3} \nabla_\mu u^\mu + \dot{E}_r \right) \chi \right] = 0.
    \label{eq:simplified transport equation 3}
\end{equation}
The first term in the round brackets describes the energy change due to adiabatic expansion or compression, while the second term accounts for radiative losses, which, at the moment, includes synchrotron losses exclusively,
\begin{equation}
    \dot{E}_r = \frac{4\sigma_T c \beta_e^2 U_B}{3 m_e^2 c^4} E^2,
\end{equation}
where $\sigma_T$ is the Thomson cross section, $\beta_e \approx 1$ is the electron velocity in units of $c$, $m_e$ is the electron mass, and $U_B = B^2/8\pi$ is the fluid rest frame magnetic energy density (in contrast with the previous Section, we do not reabsorb any factor $\sqrt{4\pi}$ in the field definition from now on). For simplicity, we neglect inverse-Compton losses, a suitable choice for sources like HBLs. 

\eq{simplified transport equation 3} is a first-order hyperbolic partial differential equation with variable coefficients and is numerically solved in our code using a second-order implicit-explicit Runge-Kutta scheme, specifically the strong stability preserving algorithm (2,2,2), described in \cite{pareschi&russo05} (see appendix \ref{sec:spectral update scheme} for more details). For self-consistency, the nonthermal particles associated with a macroparticle must remain within the computational cell that contains the macroparticle itself. This requires the maximum Larmor radius of these nonthermal particles to be smaller than the size of the computational cell. Consequently, we imposed an upper limit on the energy of the nonthermal particles \citep{vaidya+18}
\begin{equation}
    E_\mathrm{max} = \frac{e B \min(\Delta x, \Delta y, \Delta z)}{2 \beta_e},
\end{equation}  
where $e$ is the electron charge, $B$ is the laboratory-frame magnetic field interpolated at the macroparticle position, while $\Delta x$, $\Delta y$, and $\Delta z$ are the cell dimensions. This energy limit was also enforced during the shock update procedure.

The macroparticle spectra are necessary to calculate their radiative output, which corresponds solely to synchrotron emission for our purposes. For easier comprehension, in the following the quantities marked with a prime are described in the fluid rest frame, to distinguish them from the equivalent quantities defined in the laboratory frame. The synchrotron specific emissivity of a single macroparticle is
\begin{equation}
    J_\mathrm{syn}^\prime (\nu^\prime) = \frac{\sqrt{3} e^3}{4\pi m_e c^2} |\vec{B}^\prime \times \hat{\bf n}^\prime| \int \mathcal{N}^\prime(E^\prime) F\left(x\right) \, dE^\prime.
    \label{eq:syn emissivity}
\end{equation}
where $\nu^\prime$ is the fluid rest frame frequency and $\hat{\bf n}^\prime$ is the versor parallel to the line of sight. Similarly, the linearly polarized specific emissivity of a single macroparticle is given by
\begin{equation}
    J_\mathrm{pol}^\prime (\nu^\prime) = \frac{\sqrt{3} e^3}{4\pi m_e c^2} |\vec{B}^\prime \times \hat{\bf n}^\prime| \int \mathcal{N}^\prime(E^\prime) G\left(x\right) \, dE^\prime.
    \label{eq:syn pol emissivity}
\end{equation}
The two functions in the integral are defined as
\begin{gather}
F(x) = x \int_x^\infty K_{5/3}(t) dt, \\ 
G(x) = x\, K_{2/3}(x), \\ 
\text{with } x = \frac{4\pi m_e^3 c^5 \nu^\prime}{3 e E^\prime|\vec{B}^\prime \times \hat{\bf n}^\prime|},
\end{gather}
where $K_a$ is the modified Bessel function of order $a$ \citep{ginzburg&syrovatskii65}. The emissivities are expressed in the fluid rest frame, as the other quantities. However, we are interested in the emissivity in the observer frame and the RMHD simulations exclusively provide us with quantities in the laboratory frame. Therefore, first we have to transform the magnetic field and line-of-sight direction from the laboratory to the fluid rest frame,
\begin{gather}
    \hat{\bf n}' = D \left[ \hat{\bf n} + \left( \frac{\Gamma^2}{\Gamma + 1} \vec{\beta} \cdot \hat{\bf n} - \Gamma \right) \vec{\beta} \right], \\
    \vec{B}' = \frac{1}{\Gamma} \left[ \vec{B} + \frac{\Gamma^2}{\Gamma + 1} (\vec{\beta} \cdot \vec{B}) \vec{\beta} \right],
\end{gather}
where $\Gamma$ is the fluid bulk Lorentz factor and $D = [\Gamma (1-\vec{\beta}\cdot\hat{\bf n})]^{-1}$ is the relativistic Doppler factor. The emissivities are then calculated in the fluid rest frame and finally boosted to the observer frame, whose quantities are indicated with a tilde,
\begin{gather}
    \tilde{J}_{\text{syn}}(\nu) = D^2 J'_{\text{syn}}(\nu^\prime), \\ 
    \tilde{J}_{\text{pol}}(\nu) = D^2 J'_{\text{pol}}(\nu^\prime),
\end{gather}
where $\tilde{\nu} = D \nu^\prime$ is the frequency in the observer frame. This procedure is repeated for each macroparticle, and their emissivities are deposited on the fluid simulation grid using the second-order accurate triangular-shaped cloud method, a standard technique also employed in PIC simulations \citep{birdsall&langdon91}. Next, we introduce a new grid, with the $Z$ axis along the line of sight, while the $X$ and $Y$ axes on the sky plane. After interpolating the emissivities from the simulation to the new grid, the synchrotron specific intensity $I$ and flux $F$ are computed,
\begin{gather}
    \tilde{I}(\nu,X,Y,\theta) = \int \tilde{J}_\mathrm{syn}(\nu, X, Y, Z) dZ, \\ 
    \tilde{F}(\nu,\theta) = \iint \tilde{I}(\nu,X,Y) \frac{dXdY}{\mathcal{D}^2}, 
\end{gather}
where $\mathcal{D}$ is the source distance and $\theta$ is the observer viewing angle. In the end, we calculate the synchrotron polarization degree properties. For that, we need the Stokes parameters $Q$ and $U$ (circular polarization is neglected), but relativistic effects must be taken into account. Following \cite{lyutikov+03} and \cite{delzanna+06}, the two Stokes parameters on the sky plane are given by
\begin{align}
\tilde{Q} (\nu,X,Y,\theta) &= \int \tilde{J}_{\text{pol}}(\nu, X, Y, Z) \cos 2\psi \, dZ \\
\tilde{U} (\nu,X,Y,\theta) &= \int \tilde{J}_{\text{pol}}(\nu, X, Y, Z) \sin 2\psi \, dZ
\end{align}
where $\psi$ is the EVPA or polarization angle, and its sine and cosine for a specific cell are equal to
\begin{gather}
\cos 2\psi = \frac{q_X^2 - q_Y^2}{q_X^2 + q_Y^2}, \quad
\sin 2\psi = -\frac{2q_X q_Y}{q_X^2 + q_Y^2}, \quad  \text{with}  \\
q_X = (1 - \beta_Z) B_X - \beta_X B_Z, \quad
q_Y = (1 - \beta_Z) B_Y - \beta_Y B_Z, 
\end{gather}
where $\beta_{X,Y,Z}$ and $B_{X,Y,Z}$ are the interpolation of the velocity and magnetic field on the new grid. The polarization degree $\Pi$ and angle (measured clockwise from the $Y$ axis) on the sky plane as functions of the frequency are given by
\begin{gather}
    \tilde{\Pi}(\nu, X, Y, \theta) = \frac{\sqrt{\tilde{Q}(\nu, X, Y, \theta)^2 + \tilde{U}(\nu, X, Y, \theta)^2}}{\tilde{I}(\nu, X, Y, \theta)}, \\ 
    \tan 2\tilde{\psi}(\nu, X, Y, \theta) = \frac{\tilde{U}(\nu, X, Y, \theta)}{\tilde{Q}(\nu, X, Y, \theta)}.
\end{gather}
Since the source location is unresolved, we integrate the Stokes parameters over the sky plane, obtaining the polarization degree and angle solely as functions of frequency.

The update of macroparticle spectra at shocks is based on the work of \cite{vaidya+18} and \cite{mukherjee+21}. First, we implemented the shock-detection algorithm proposed in \cite{mignone12}. We flag as shock the simulation cells where these two conditions are satisfied:
\begin{equation}
    \nabla \cdot \vec{v} < 0 \quad \text{and} \quad \frac{|{\nabla P}|}{P} > \zeta,  
    \label{eq:shock detection}
\end{equation}
where $\zeta$ is an arbitrary threshold. When a macroparticle enters a flagged zone, the spectrum is not updated, and the pressure interpolated at each step is recorded. Once the macroparticle exits the flagged zone, the step with the lowest pressure is identified as ``upstream,'' while the step with the highest pressure is designated as ``downstream.'' The spectrum of the exiting macroparticle is then updated using the fluid quantities interpolated at upstream and downstream points. In \cite{vaidya+18}, the post-shock spectrum is reset to a power-law distribution, while its normalization and minimum energy are computed by requiring the macroparticle to have a fraction of the fluid energy and number density.

This approach presents two drawbacks: first, the macroparticle history, namely its spectrum temporal evolution before the shock, is forgotten, second, a computational cell could contain more than one shocked macroparticle and even non-shocked macroparticles; therefore, fluid energy and number density must be distributed among the shocked macroparticles taking into account these effects. We adopt the solution proposed in \cite{mukherjee+21}. Consider a particle entering the shock with spectrum $\mathcal{N}^\prime_\mathrm{pre}$, the post-shock spectrum $\mathcal{N}^\prime_\mathrm{post}$ is updated through a convolution
\begin{equation}
    \mathcal{N}^\prime_{\text{post}}(E') = C \int_{E'_{\text{min}}}^{E'} \mathcal{N}^\prime_{\text{pre}}(E'') \left( \frac{E'}{E''} \right)^{-q+2} \frac{\mathrm{d}E''}{E''}.
    \label{eq:convolution update}
\end{equation}
The upstream and downstream conditions determine the index $q$ (see later), while the minimum energy $E'_\mathrm{min}$ is fixed to the minimum energy of the pre-shock spectrum. With the convolution, the macroparticle history is not neglected. Specifically, in case a macroparticle encounters a strong shock followed by a weak shock, the spectrum is not anomalously steepened.

The normalization constant $C$ is determined following a two-step procedure. First, we calculate the internal energy density to distribute among the shocked macroparticles
\begin{equation}
    \Delta E = f_E \rho \epsilon - \sum_\text{nsh} \varepsilon, 
\end{equation}
where $\rho \epsilon$ is the fluid internal energy density, $f_E$ is the fraction of the internal energy density available to the shocked macroparticles, and $\sum_\text{nsh} \varepsilon$ is the sum of the energy densities of the non-shocked macroparticles. Note that $\epsilon$ is the fluid internal energy per unit of mass and it is determined by the Taub-Matthews equation of state since we are considering a relativistic fluid \citep{mignone07}. If $\Delta E < 0$, the shocked macroparticle spectra are not updated. Second, we compute the fluid number density to distribute among the shocked macroparticles
\begin{equation}
    \Delta N = f_N \frac{\rho}{\mu m_a} - \sum_{\text{sh \& nsh}} N_i,
\end{equation}
where $m_a$ is the atomic mass unit, $\mu$ is the mean molecular weight for ionized gas, $f_N$ is the fraction of the fluid number density available to the shocked macroparticles, and $ \sum_{\text{sh \& nsh}} N_i$ is the sum of the number densities of non-shocked macroparticles along with the contribution of the shocked macroparticles, normalized in the previous step. If $\Delta N < 0$, the normalization constants $C$ are reduced to obtain $\Delta N = 0$. 

\subsection{Upstream and wall frame}

In our framework, we aimed to refine the shock acceleration approach proposed in \cite{vaidya+18}. We calculated the shock quantities at MHD scales (e.g., the upstream and downstream magnetic fields) to derive quantities (e.g., upstream magnetization) that can be directly compared with PIC simulation results. In their turn, kinetic simulations provide key features of the post-shock spectrum, including the spectral index $q$, the maximum Lorentz factor $\gamma_\mathrm{max}$, and the fractions of internal energy and particle number density available to shocked macroparticles, denoted as $f_E$ and $f_N$, respectively. 

However, not all shock configurations lead to efficient nonthermal particle acceleration. Given moderate magnetization and a relativistic shock, the upstream magnetic obliquity $\theta_\mathrm{up}$ (i.e., the angle between the upstream magnetic field and the shock normal) plays a key role. In fact, in a magnetized medium, gyrating particles are constrained to move along the magnetic field lines, which in turn are advected downstream by the flow. Therefore, if the upstream magnetic obliquity is too large, the particles should move faster than the speed of light to return upstream \citep{begelman&kirk90}. This scenario is called superluminal shock, and kinetic simulations confirm that in this case, no acceleration takes place (\citealt{crumley+19} for the transrelativistic case, \citealt{sironi&spitkovsky09, sironi&spitkovsky11} for the relativistic case). More quantitatively, a shock is defined as superluminal when $\cos \theta_\mathrm{up} < \beta_\mathrm{sh,up}$, where $\beta_\mathrm{sh,up}$ is the shock speed normalized to $c$ \citep{kirk&heavens89, begelman&kirk90}. Both quantities are defined in the frame where the upstream is at rest. Only subluminal shocks can efficiently accelerate particles. PIC simulations show that in this scenario the post-shock spectrum displays a nonthermal power-law tail (\citealt{crumley+19} for the transrelativistic case, \citealt{sironi&spitkovsky09, sironi&spitkovsky11} for the relativistic case). Ultimately, when a macroparticle crosses a shock, it is essential to compute the upstream magnetic inclination and shock speed. We need to transform the fluid quantities from the laboratory frame to the upstream rest frame, where the superluminal condition is defined.

Moreover, PIC simulations show that the features of the post-shock spectrum depend on the upstream magnetization, magnetic inclination, and Lorentz factor (\citealt{crumley+19} for the transrelativistic case, \citealt{sironi&spitkovsky09, sironi&spitkovsky11} for the relativistic case). However, in PIC simulations the fluid quantities are computed in the wall frame, defined as the frame in which the normal component of the downstream velocity is null. Therefore, it is also necessary to transform the fluid quantities from the laboratory to the wall frame to take advantage of the results of the PIC simulations. For the implementation of the transformations from the laboratory to the upstream and wall frames, refer to appendix \ref{sec:frame transformations}.  

\subsection{Cooling versus advection time}

Before showing our results, we first checked the consistency of our framework. We are implicitly assuming that the shock structures observed in PIC simulations are much smaller than the scales we consider. For fluid quantities, this assumption is naturally satisfied, as their temporal and spatial evolution is described by the RMHD equations, which, by definition, treat the plasma on scales much larger than shock kinetic scales. Conversely, the situation is less clear for the scales governing the evolution of the macroparticle spectrum. In particular, the synchrotron cooling length of the nonthermal particle can be comparable with the decay length of the downstream microturbulence. In this scenario, acceleration and cooling would not be fully spatially separable since the contribution of the downstream region close to the shock surface to the cooling process is not negligible. A more quantitative evaluation is thus necessary (see \citealt{vanthieghem+20} for a similar discussion). 

PIC simulations show that the downstream microturbulence decay length $\ell$ is on the order of $10^3 c/\omega_{p,i}$, where $\omega_{p,i}$ is the ion plasma skin depth. From fluid simulations, the thermal proton number density inside the jet is equal to $n_{p,th} = 1 \unit{cm^{-3}}$. Hence, we obtain $\ell \sim 10^{10} \unit{cm}$. Assuming a strong shock, the advection velocity in the downstream region is $v_\mathrm{adv} \approx c/3$, so the time a particle spends in the microturbulence before being advected in the far downstream is $t_\mathrm{adv} \sim 3\ell/c$. The synchrotron cooling time $t_\mathrm{cool}$ immediately after the shock depends on the microturbulence magnetic field strength $\delta B$. From our fluid simulations, we can estimate that the downstream magnetic field in the fluid frame is $B_2^\prime \sim 5 \times 10^{-2} \unit{G}$. However, PIC simulations show that the ratio between the microturbulence and far downstream magnetic energy density can be significantly large ($\sim 50$, e.g., \citealt{tavecchio+18}), which implies that $\delta B \sim 10^{-1} \unit{G}$. The comparison between the two timescales gives an upper limit for the Lorentz factor below which the microturbulence contribution to cooling is negligible:
\begin{equation}
    \gamma < \frac{2\pi m_e c^2}{\sigma_T B^2 \ell} \approx 7.7 \times 10^{10} \left(\frac{\delta B}{10^{-1} \unit{G}} \right)^{-2} \left ( \frac{\ell}{10^{10} \unit{cm}} \right)^{-1}.
    \label{eq:cooling vs decay}
\end{equation}
The hypothesis that the cooling in the microturbulence region is negligible is valid in the Lorentz factor range of interest. Assuming the magnetic field in the emitting region $B_2^\prime \sim 5 \times 10^{-2} \unit{G}$ and a relativistic Doppler factor $D \sim 10$, the Lorentz factor of the electrons emitting in the keV emission or lower is $\lesssim 4 \times 10^5$, well below the limit given by \eq{cooling vs decay}.

\section{Results}
\label{sec:results}

In this section, we report the results of our post-processing framework, applied to simulations of recollimated magnetized jets. 

\begin{figure}[htbp]
\centering
\includegraphics[width=\columnwidth]{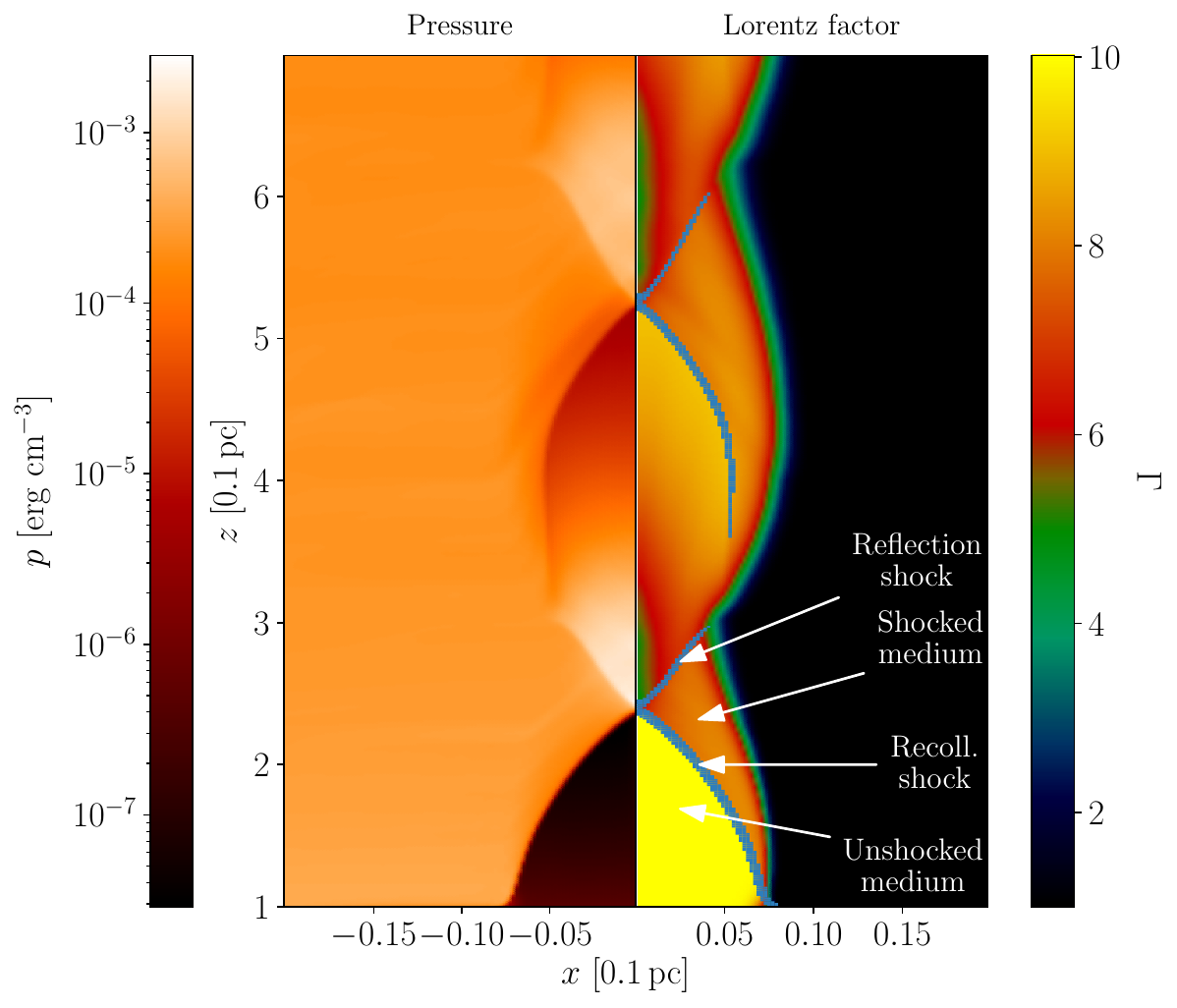}
\caption{Map of the pressure (left) and Lorentz factor (right) in the $xz$ plane (Simulation A). The main structures are indicated. The cells tagged as shock regions in the $xz$ plane are highlighted in light blue.}
\label{fig:dual plot}
\end{figure}

Our post-processing procedure was applied to the axisymmetric stationary state, obtained from 2D simulations and interpolated to 3D. The steady state corresponds to a jet collimated by a higher-pressured environment, as shown in Fig. \ref{fig:dual plot}. Since the collimated jet is supersonic, a recollimation shock forms \citep{komissarov&falle97}. After this shock, the pressure increases, reaching an equilibrium with the external medium. Meanwhile, the Lorentz factor decreases, but because the recollimation shock is oblique, the flow remains relativistic. As streamlines converge toward the jet axis, they are reflected by the reflection shock. Because the incoming flow is nearly perpendicular to the reflection shock, the dissipation of kinetic energy is more significant at this stage. This results in a further pressure increase, causing the downstream region of the reflection shock to become over-pressured compared to the post-recollimation shock region and the external medium. Consequently, the jet re-expands. However, if the external medium maintains a stationary pressure profile, the jet undergoes another recollimation, leading to an oscillatory behavior of the jet radius. In this process, the relativistic fluid repeatedly experiences cycles of recollimation and reflection shocks. Bulk kinetic energy is cyclically converted to thermal energy through shock heating and then reconverted to kinetic energy via adiabatic cooling \citep{matsumoto+12}. The confinement of jets, leading to the development of the chain of recollimation-reflection shocks, has been extensively studied using 2D relativistic hydrodynamic (RHD) simulations (e.g., \citealt{gomez+95, agudo+01, bodo&tavecchio18}) and as well as a few 2D RMHD simulations \citep{mizuno+15, marti+16}, which (at low-magnetization, as in our case) provide similar results to RHD ones. Our 2D RMHD simulations are in agreement with previous results. However, jets often exhibit more complex behavior due to various instabilities that depend on the jet intrinsic properties \citep{birkinshaw96, bodo+13, bodo+19, gourgouliatos&komissarov18} and its surrounding environment \citep{porth&komissarov15, tchekhovskoy&bromberg16}. A full 3D simulation is required to study recollimation instabilities in detail, but such an analysis is beyond the scope of this paper (Boula et al. in prep.).

Below, we outline the selected values for the free parameters used in the post-processing code. We fixed $\zeta = 1$ (see \eq{shock detection}), which is smaller than previously used values (for example, \citealt{mignone07}). After some tests, we observed that large values of $\zeta$ produce thin shock regions, especially at the reflection shock. By selecting a smaller value, we obtained thicker shock regions that allow a more accurate determination of upstream and downstream states, leading to improved estimates of the derived quantities, such as the shock normal and speed. 

We must specify the line of sight direction, which is crucial for computing the relativistic Doppler factor. The viewing direction is defined by the polar $\theta$ (also called in the text observer viewing angle) and azimuthal angle $\phi$. Due to the axisymmetric nature of our solutions, the azimuthal angle is irrelevant, so, without loss of generality, we set $\phi = 0$. As for the polar angle, we fixed $\theta = 5^\circ$, suitable for blazars. For a discussion on the dependence of the polarimetric properties on the observing viewing angle, see subsection \ref{subsec:turning on acceleration}.

Macroparticles were injected into the jet base at the start of the post-processing. Since we were considering a steady-state scenario, continuously injecting macroparticles over multiple simulation time steps is equivalent to an instantaneous injection at a single simulation time step. In this context, the full history of a macroparticle moving along the jet provides for macroparticles injected at later times. Two macroparticles were injected into each jet base cell and we selected the time step to ensure two macroparticles for every step in the $z$-direction, namely $\Delta t = 0.005\,z_0/c$ in code units. We fixed the maximum time $t_\mathrm{max} = 550 \Delta t = 2.75\,z_0/c$. This ensures that the motion of most of the macroparticles ends before reaching the second recollimation shock, which was intentionally excluded from our analysis (see later).

Each macroparticle spectrum was sampled on an energy grid spanning from $E_\mathrm{min} = 3 m_e c^2$ to $E_\mathrm{max} = 10^9 m_e c^2$. The energy grid is logarithmically spaced, with a resolution of 5 points per decade. Regarding the initial energy distribution of macroparticles, we opted for a steep power law $\mathcal{N} \propto E^{-\alpha}$ with $\alpha = 9$, and limited the Lorentz factor range to $\gamma \in [10^1, 10^2]$. With this injection spectrum, the macroparticle history before the first shock is irrelevant. Even if shock conditions are not optimal for strong acceleration, the convolution in \eq{convolution update} assures that a slightly flatter index dominates, resulting in a power law with the just-mentioned index, effectively erasing the influence of the steep injection spectrum. The values chosen for the free parameters are summarized in Tab. \ref{tab:free parameters}. 

\begin{table}[t]
\centering
\caption{Recap free parameter table, indicating their names, symbols, and values}
\resizebox{\columnwidth}{!}{
\begin{tabular}{lc}
\hline
Parameter                                                   & Value  \\ \hline
Shock detection pressure threshold ($\zeta$)                & $1$           \\
Viewing angle ($\theta$)                                    & $5^\circ$         \\
Macroparticle injection density                             & $2 \unit{macroparticles/cell}$    \\ 
Time step ($\Delta t$)                                      & $0.005\,z_0/c$    \\ 
Maximum time ($t_\mathrm{max}$)                             & $2.75\,z_0/c$    \\ 
Energy grid range  ($[E_\mathrm{min}, E_\mathrm{max}]$)     & $[3 m_e c^2,10^9 m_e c^2]$ \\
Energy grid resolution                                      & $5 \unit{points/decade}$ \\
Injection power index ($\alpha$)                            & $9$    \\ 
Injection Lorentz factor range                              & $[10^1, 10^2]$    \\ \hline 
\end{tabular}}
\label{tab:free parameters}
\end{table}

We did not consider the entire simulation domain for our post-processing analysis. Instead, we focused on a reduced region defined by $[-0.2, 0.2]\times[-0.2, 0.2]\times[1, 4.5]$, in units of $z_0$. This choice allowed us to exclude most of the external medium, where no shocks, and therefore acceleration, occur. Moreover, we were exclusively interested in the first couple of recollimation-reflection shocks. As shown by \citet{costa23} in unmagnetized recollimated jets, the turbulence generated downstream of the reflection shock dissipates the bulk kinetic energy of the flow and it weakens the subsequent shocks, to finally suppress their formation. Moreover, we anticipated that the reflection shock is the site where most high-energy radiation is produced (see later). Since we expected a similar behavior for jets with low magnetization, we left out shocks beyond the first reflection shock in our analysis. 

\begin{figure}[htbp]
\centering
\includegraphics[width=\columnwidth]{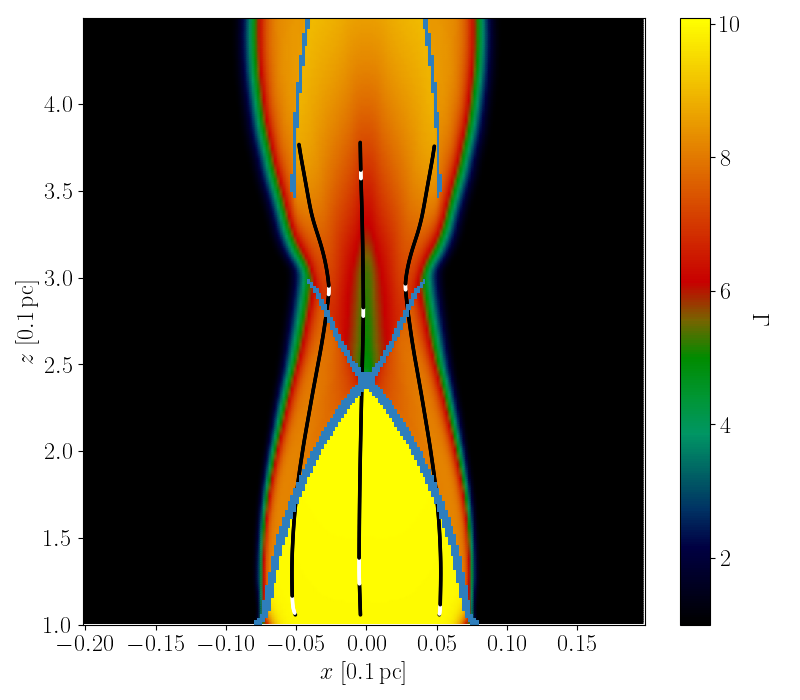}
\caption{Map of the Lorentz factor in the $xz$ plane with the projection of the trajectory of 3 macroparticles (number 799, 3983, 7789, Simulation C). The trajectory color represents the macroparticle status: white when inside a shock region and black when outside. The cells tagged as shock regions in the $xz$ plane are highlighted in light blue.}
\label{fig:macros moving injet}
\end{figure}

In \fig{macros moving injet}, we present a 2D slice at $y=0$ showing the Lorentz factor distribution of Simulation C. Overlaid on this, we display the trajectories of three randomly selected macroparticles. These macroparticles originate at the jet base and, after a few steps, encounter the first recollimation shock. Since the macroparticles move in three dimensions, their exact time steps of being tagged as ``shocked'' do not necessarily align with the shock region in the $xz$ plane. However, if we imagine rotating the $xz$-shock structure around the $z$-axis, we obtain the full 3D shock region. This visualization clarifies that the macroparticles do indeed traverse the shock. Beyond the recollimation shock, the jet narrows, causing the macroparticles' trajectories to shift closer to the $z$-axis as they follow the jet streamlines. Upon reaching the first reflection shock, their paths deviate outward again due to the jet's re-expansion. Finally, only the central macroparticle interacts with the second recollimation shock. As already pointed out, the post-processing duration was selected to ensure that only a small fraction of macroparticles reach the second recollimation shock. 

\subsection{Subluminal versus superluminal scenario}

\begin{figure*}[htbp]
\centering
\includegraphics[width=.97\textwidth]{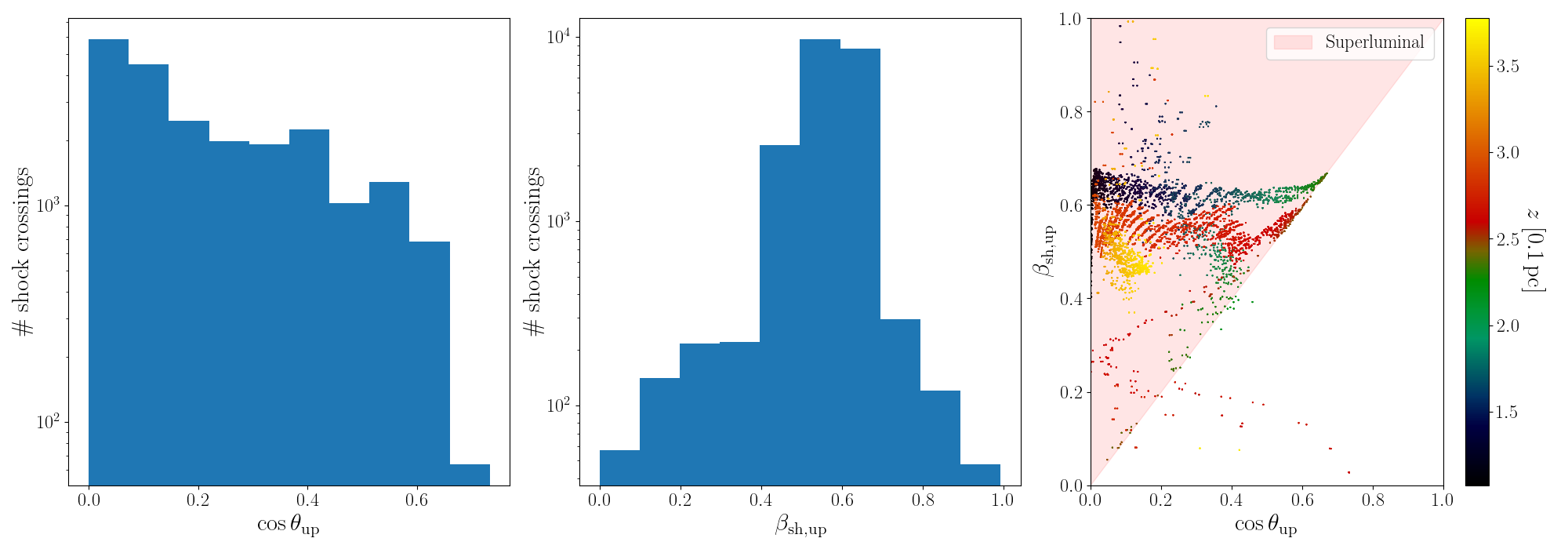}
\caption{Left and middle panels: Distribution of the upstream magnetic inclination and shock speed in the upstream rest frame for Simulation A. Right panel: Comparison of these two quantities for each configuration. The dot color indicates the $z$ coordinate of the macroparticle exiting point from the shock. The configurations inside the red region are superluminal.}
\label{fig:super vs sub A}
\end{figure*}

Every time a macroparticle crossed a shock, the first step was to identify whether the configuration is subliminal or superluminal. In the superluminal scenario, no particle acceleration occurs, and therefore the post-shock spectrum remains unchanged. Moreover, we computed the wall frame parameters required for spectrum updates, specifically the upstream magnetic field inclination, magnetization, and Lorentz factor, so that these values are readily available if needed. By comparing these quantities with PIC simulations, we could determine the slope of the post-shock energy distribution for subluminal cases. Figure \ref{fig:super vs sub A} illustrates the distribution of the essential parameters used to classify shock crossings in Simulation A (refer to appendix \ref{sec:versus B and C} for Simulation B and C). In addition, we plot each shock crossing in the parameter space defined by $\cos \theta_\mathrm{up}$ and $\beta_\mathrm{sh, up}$. This allows us to visually distinguish which configurations are superluminal or subluminal. 

Except for a few outliers, likely due to inaccuracies in reconstructing the upstream rest frame parameters, almost all configurations are superluminal in all scenarios. For Simulation A and B, at low $z$ (blue points) recollimation shock crossings lie deep within the superluminal region, and as the crossing occurs at higher $z$ (green points), the parameters gradually move closer to the boundary between superluminal and subluminal regimes. Near the top of the recollimation shock, some configurations even become subluminal in Simulation A (dark green points). This behavior can be explained by the relative inclination between the shock normal and the streamlines. The inclination is high for low $z$, resulting in full superluminal configurations. As $z$ grows, the inclination decreases, leading to shock crossing near the superluminal-subluminal limit. The reflection shock behaves oppositely. At low $z$, the shock crossings remain close to the superluminal-subluminal boundary, with a few subluminal configurations in Simulation A (dark red points), but as $z$ increases, the shock crossings revert to fully superluminal conditions (red points). Overall, the recollimation shock transitions from strongly superluminal behavior toward the subluminal threshold with increasing $z$, while the reflection shock moves from near-sub to firmly superluminal conditions as $z$ grows. Compared to the recollimation shock, the reflection shock speed is lower, a trend also evident for the few macroparticles that encounter the second recollimation shock (yellow points). In Simulation C, all configurations exhibit a strong superluminal behavior. This is primarily due to the magnetic field configuration, as the field is predominantly toroidal. Consequently, the upstream magnetic inclination remains low in nearly all cases.

Therefore, we conclude that, assuming a laminar flow, no particle acceleration can occur in the shocks associated with recollimation.

\subsection{Turning on acceleration}
\label{subsec:turning on acceleration}

No acceleration is possible, except for very few macroparticles. Thus, we decided to change our starting assumptions. Previously, we were supposing that the fluid is laminar on all lengths, from fluid to kinetic scales. The distinction between superluminal and subluminal configuration is crucial if the upstream flow is laminar (\citealt{crumley+19} for the transrelativistic case, \citealt{sironi&spitkovsky09, sironi&spitkovsky11} for the relativistic case). Recent simulations indicate that the presence of turbulence or inhomogeneities on kinetic scales in the upstream medium can reignite acceleration, even in magnetized and superluminal configurations. In one scenario, turbulence frees particles from tightly following magnetic field lines, preventing their advection into the downstream region and thus enabling acceleration at the shock \citep{bresci+23}. In another scenario, the shock interaction with these inhomogeneities favors the formation of intense downstream turbulence, which accelerates particles, leading to the formation of a power-law tail in the post-shock spectra \citep{demidem+23}. For both scenarios, the parameter space is not extensively explored, as for laminar flows. Therefore, we could not compare the shock crossing parameters directly with the simulations to deduce the post-shock properties. For this reason, we were forced to use phenomenological formulae. 

For the power law index, we opted for a generalization of the classical formula \citep{drury83}, which was originally derived under the assumption of isotropic diffusion in the particle velocity angle relative to the shock normal \citep{keshet&waxman05},
\begin{equation}
    q = \frac{3\beta_{1,\mathrm{NIF}} - 2\beta_{1,\mathrm{NIF}}\beta_{2,\mathrm{NIF}}^2 + \beta_{2,\mathrm{NIF}}^3}{\beta_{1,\mathrm{NIF}} - \beta_{2,\mathrm{NIF}}} - 2,
    \label{eq:relativistic shock slope}
\end{equation}
where $\beta_{1,\mathrm{NIF}}$ and $\beta_{2,\mathrm{NIF}}$ are respectively the upstream and downstream velocity normalized to $c$ in the normal incident frame (NIF), where the shock is at rest and the upstream velocity is directed along the shock normal (for more details about the transformation from the laboratory frame to the NIF, see appendix \ref{sec:frame transformations}). This formula is justified if strong upstream turbulence detaches particles from magnetic field lines, like in the scenario investigated in \cite{bresci+23}, effectively making the shock behave as if it were low magnetized, as assumed in \cite{keshet&waxman05}. 

Regarding the maximum energy of the post-shock spectra, we followed the standard approach (e.g., \citealt{kirk94,boettcher&dermer10,baring17}). Supposing that the acceleration time is a multiple of the gyrofrequency (i.e., $t_a = \lambda 2 \pi r_g/c$; this corresponds to assume a ``gyro-Bohm'' spatial diffusion coefficient, e.g., \citealt{kirk94}) and the main cooling process is synchrotron emission, the maximum Lorentz factor, dictated by equilibrium between energy gains and losses, is given by
\begin{equation}
    \gamma_\mathrm{max} = \frac{E_\mathrm{max}}{m_e c^2} = \left( \frac{9 m_e^2 c^4}{8\pi B^\prime \lambda e^3}\right)^\frac{1}{2},
    \label{eq:maximum gamma}
\end{equation}
where $B^\prime$ is the fluid rest frame magnetic field interpolated at the point where the macroparticle exits the shock region. The free parameter $\lambda$, also called acceleration efficiency, contains the shock microphysics and can be interpreted as the ratio between the diffusion mean free path and the particle gyroradius. We set $\lambda = 1000$, while in \cite{vaidya+18} this parameter is derived through a more complex approach, which gives, on average, $\lambda \sim 1$, namely the so-called Bohm limit. As shown in \cite{garson+10, boettcher+12, summerlin&baring12}, this value implies extremely efficient shocks, resulting in electrons producing synchrotron radiation up to the MeV band. Thus, we adopted a significantly larger $\lambda$, ensuring that synchrotron emission peaks inside the X-ray band, as observed for HBLs and EHBLs (in appendix \ref{sec:sim fluxes} we report the flux as a function of frequency for the three cases). On kinetic scales, this value corresponds to relatively weak microturbulence around the shock, resulting in a slower diffusion and, thus, a higher acceleration time. This is in tension with the previously described scenario, where particles are either detached by magnetic field lines or directly accelerated by strong turbulence. However, the maximum particle energy could still be limited by the spatial scale of the microturbulence. As shown in \cite{sironi+13}, if a particle gyroradius exceeds the scale of the microturbulence, it will be advected away from the shock front, effectively halting the acceleration process\footnote{Applying Eq. 16 of \cite{sironi+13} we checked that a likely upper limit for the electron Lorentz factor is $\gamma_{\rm max}\approx 10^6$.}. This mechanism could alleviate the discrepancy between the required acceleration efficiency for effective particle acceleration and the observed synchrotron emission limits. While this effect has been verified in PIC simulations of laminar flows, it has not yet been fully explored in turbulent or inhomogeneous flow conditions; therefore, we stick to \eq{maximum gamma}.

For a laminar flow, the ratio of nonthermal to thermal energy density is $10^{-3}$ for a transrelativistic shock \citep{crumley+19} and $10^{-1}$ for a relativistic shock \citep{sironi&spitkovsky09, sironi&spitkovsky11}. The ratio of nonthermal to thermal number density is only available for the relativistic scenario, where it is $5 \times 10^{-2}$ \citep{sironi&spitkovsky09, sironi&spitkovsky11}. In the case of turbulent and inhomogeneous scenarios, these values are unavailable due to the limited runtime of the simulations. Given this, we adopt intermediate values of $f_N = f_E = 10^{-2}$. However, it is important to note that since we are primarily concerned with intensity ratios, these specific values do not affect the final results.

\begin{figure}[htbp]
\centering
\includegraphics[width=\columnwidth]{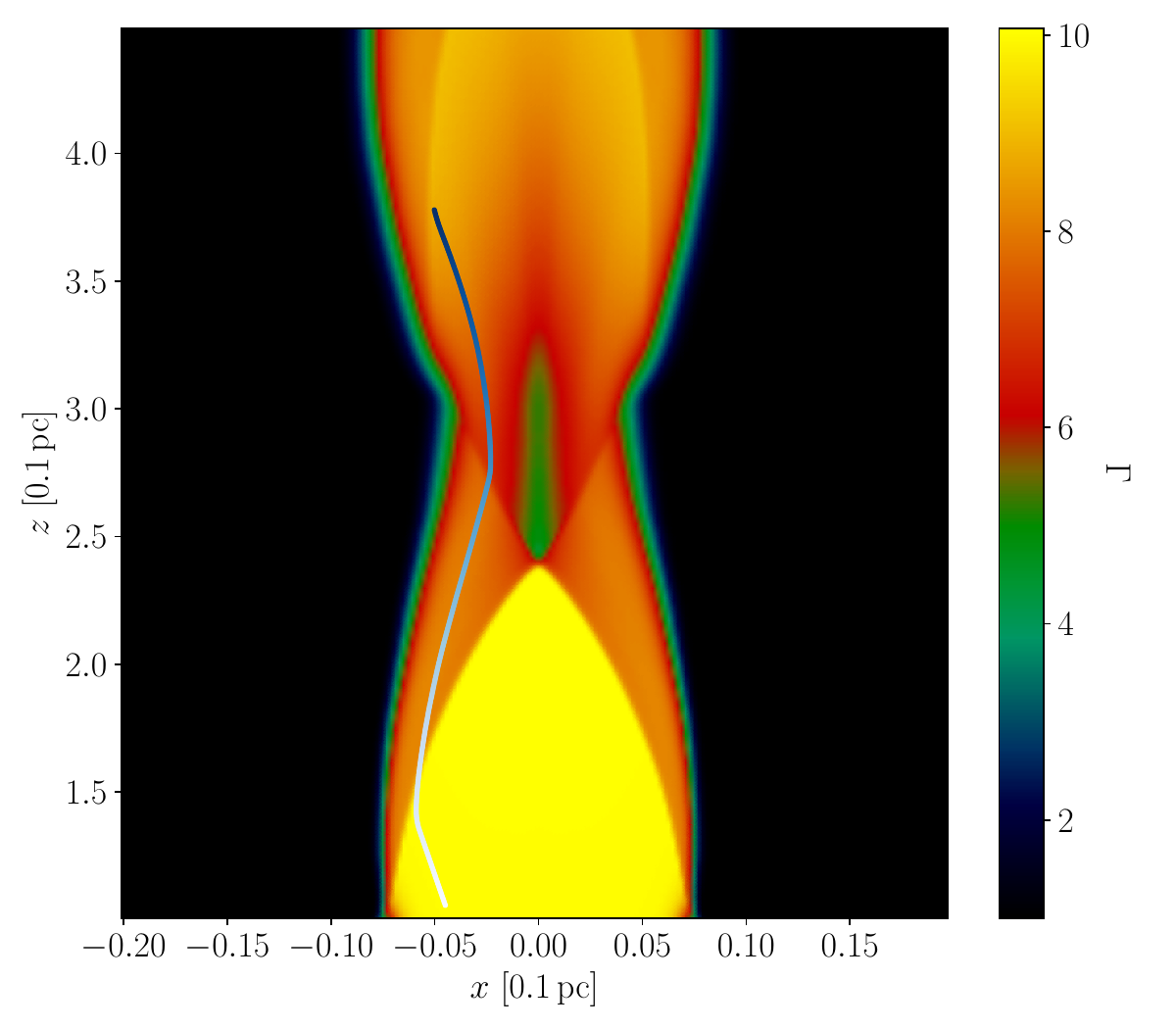}
\caption{Map of the Lorentz factor in the $xz$ plane with the projection of the trajectory of a macroparticle (number 1135, Simulation B). The color of the trajectory represents the time step, see \fig{macro spectrum}}
\label{fig:macro moving injet}
\end{figure}

\begin{figure}[htbp]
\centering
\includegraphics[width=\columnwidth]{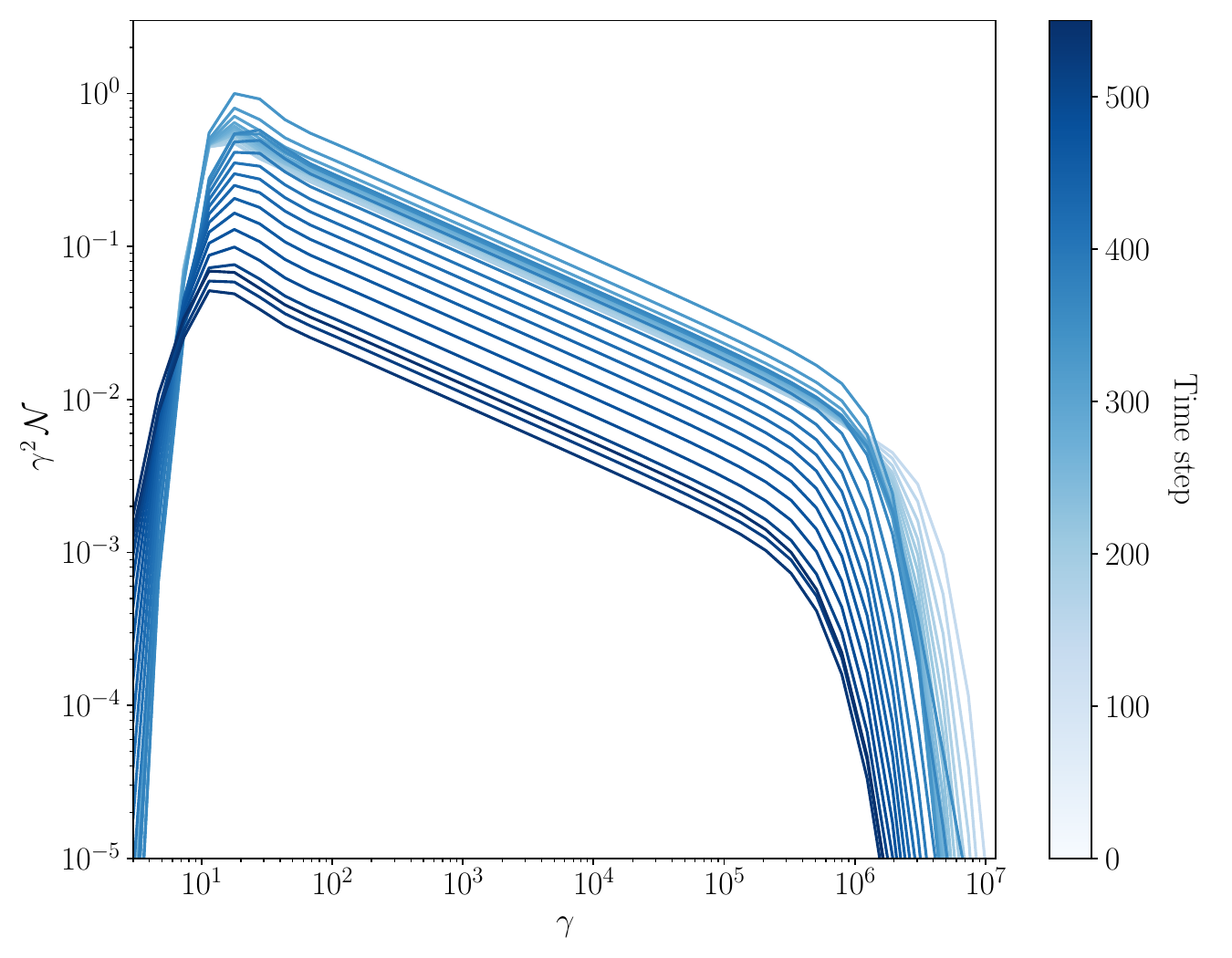}
\caption{Time evolution of the normalized spectrum of the macroparticle in \fig{macro moving injet}}
\label{fig:macro spectrum}
\end{figure}

Figs. \ref{fig:macro moving injet} and \ref{fig:macro spectrum} represent a macroparticle trajectory and its spectral time evolution. The macroparticle is injected into the jet base, then it is advected by the flow until it reaches the recollimation shock. Before that, the spectrum is not visible in \fig{macro spectrum} because of the low normalization. Thanks to shock acceleration, the macroparticle spectrum becomes a power law, but, as it is advected from the recollimation to the reflection shock, an exponential cut-off emerges for high Lorentz factors due to the synchrotron cooling. When the macroparticle encounters the reflection shock, it is accelerated again. Since the magnetic field is higher due to the double shock compression, the maximum Lorentz factor is lower. On the other side, the internal energy and particle number density increase after both shocks, resulting in a larger normalization for the post-reflection shock spectrum. Finally, the macroparticle is again advected by the flow and cooled down due to synchrotron and, to a smaller extent, adiabatic losses.       

\begin{figure}[htbp]
\centering
\includegraphics[width=\columnwidth]{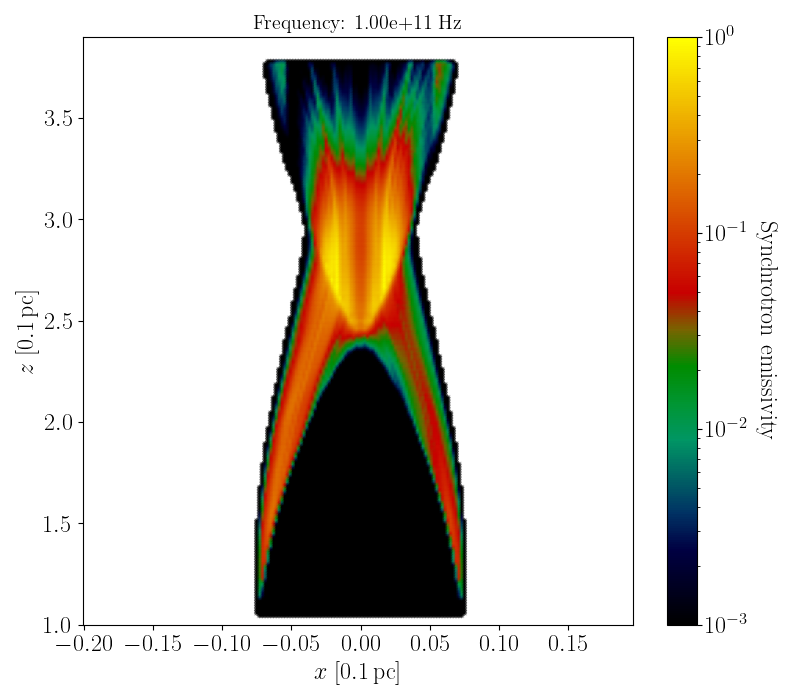}
\caption{Normalized observer frame synchrotron emissivity in the radio band of a 10-code units slab centered on $xz$ plane (Simulation B).}
\label{fig:radio emissivity}
\end{figure}

\begin{figure}[htbp]
\centering
\includegraphics[width=\columnwidth]{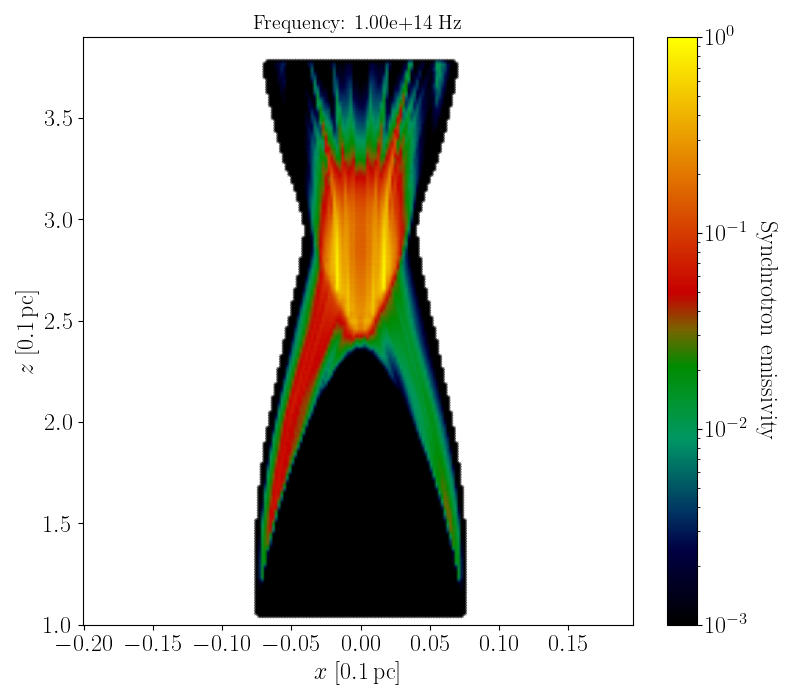}
\caption{Same as Fig. \ref{fig:radio emissivity} but in the optical band.}
\label{fig:optical emissivity}
\end{figure}

\begin{figure}[htbp]
\centering
\includegraphics[width=\columnwidth]{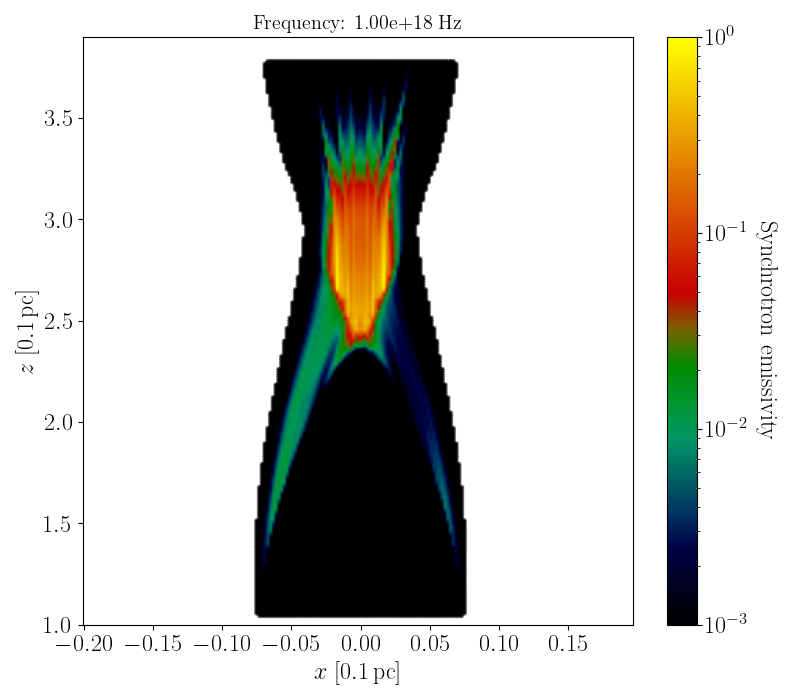}
\caption{Same as Fig. \ref{fig:radio emissivity} but in the X-ray band.}
\label{fig:xray emissivity}
\end{figure}

\begin{figure}[htp]
\centering
\includegraphics[width=\columnwidth]{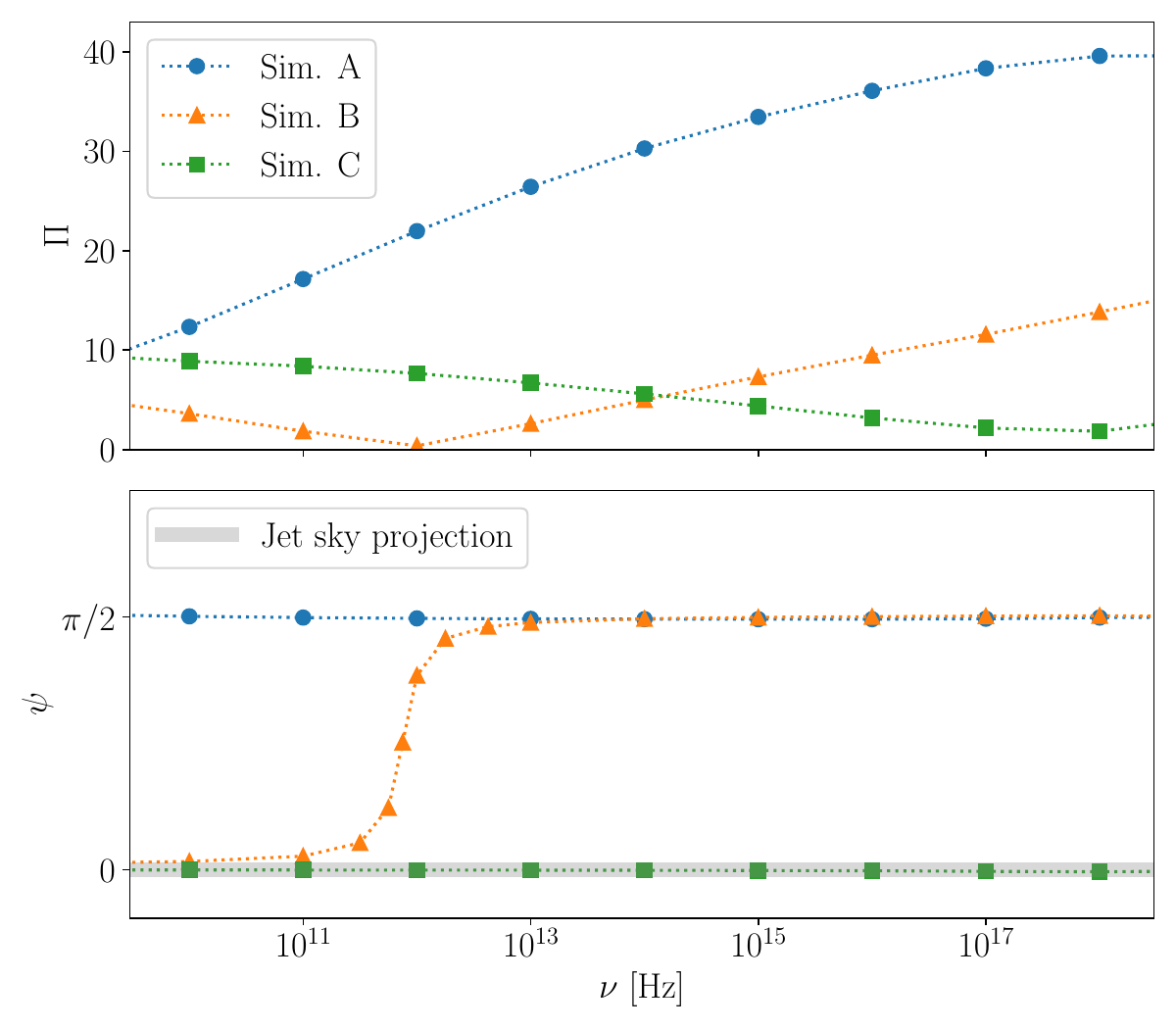}
\caption{Polarization degree (top panel) and angle (bottom panel) of Simulation A (blue circles), B (orange triangles), and C (green squares). The polarization angles are measured clockwise from the projection of the jet axis onto the sky plane (solid gray line).}
\label{fig:polarization features}
\end{figure}

Figs. \ref{fig:radio emissivity}, \ref{fig:optical emissivity}, and \ref{fig:xray emissivity} show the emissivity in the observer frame for Simulation B of a slice centered on the $xz$ plane in the radio, optical, and X-ray band. The emissivities of Simulation A and C present similar features (see Appendices \ref{sec:sim A emissivities} and \ref{sec:sim C emissivities}). The evident asymmetry of the emissivity is due to the combined effect of relativistic beaming and the $|\vec{B}^\prime \times \hat{\bf n}^\prime|$ term in the synchrotron formulae, see Eqs. \eqref{eq:syn emissivity} and \eqref{eq:syn pol emissivity}. The emission from portions of the flow with streamlines closely oriented toward the observer (e.g., the left side of the recollimation shock downstream) is more amplified, compared to the emission associated with the plasma whose velocity is oriented in other directions. 

We neglected Faraday rotation and synchrotron self-absorption, which are significant below $10^{11} \unit{Hz}$ \citep{tavecchio10}. Consequently, our model is not suitable for accurately reproducing radio emission from regions at parsec-scale distances. However, the observed radio emission is primarily produced by regions located farther from the supermassive black hole.

\begin{figure}[htp]
\centering
\includegraphics[width=\columnwidth]{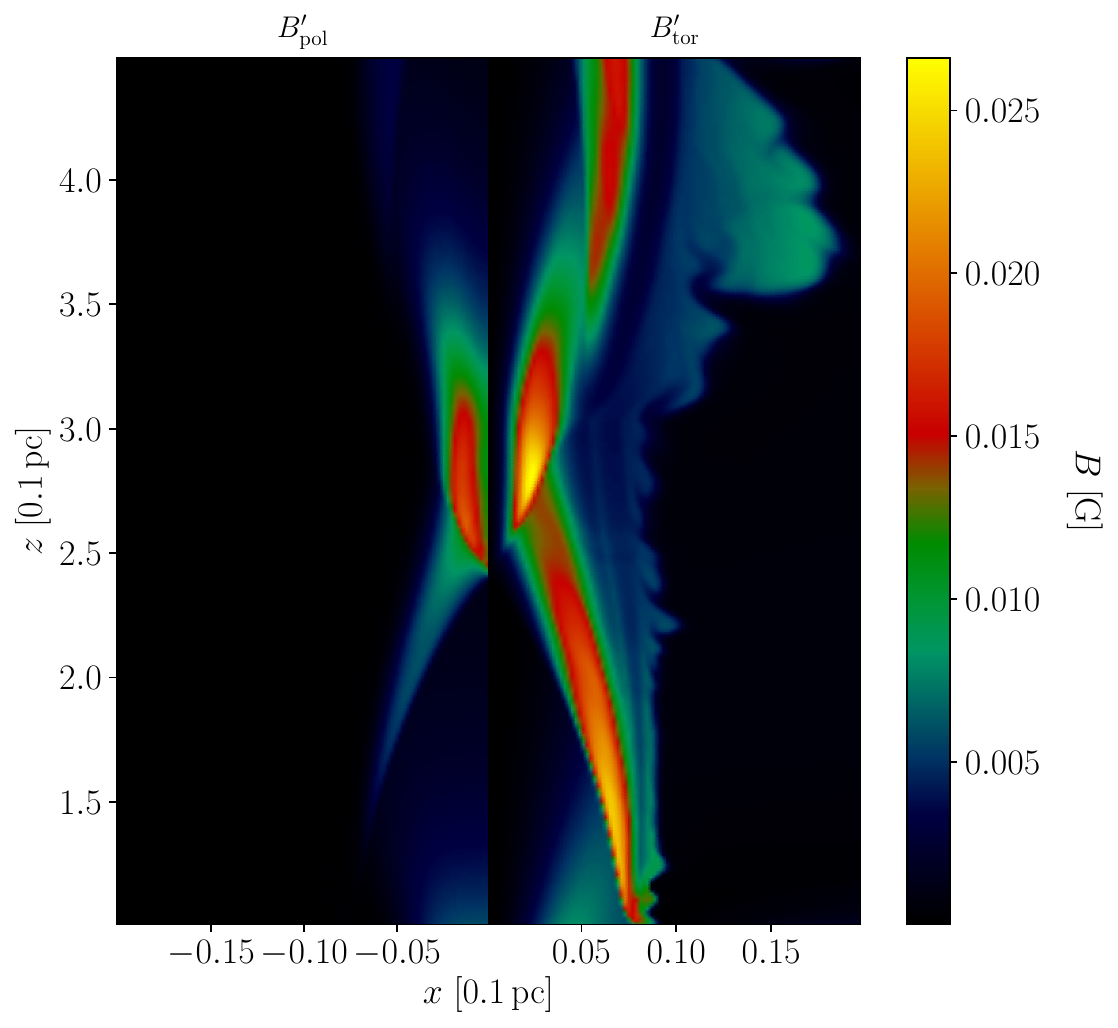}
\caption{Map of the poloidal (left panel) and toroidal (right panel) component of the fluid rest frame magnetic field in the $xz$ plane (Simulation B).}
\label{fig:poloidal vs toroidal magnetic field}
\end{figure}

\fig{polarization features} displays the polarization parameters of the three configurations. In all cases, we obtain a strong chromaticity. In Simulations A and B, the polarization degree increases with the frequency, as observed by multiwavelength polarization campaigns of HBLs and EHBLs in a quiescent state. In particular, the polarization degree values of Simulation B are comparable with the IXPE and multiwavelength results. The chromaticity can be traced back to the structure of the synchrotron emissivity and the magnetic fields. In the radio band (Fig. \ref{fig:radio emissivity}), the region where the emissivity is significant (from now on emission region) includes both the downstream regions of the recollimation and the reflection shocks. In these regions, both components of the field, toroidal and poloidal, are significant (Fig. \ref{fig:poloidal vs toroidal magnetic field}), thus reducing the total polarized fraction of the emission. On the other hand, the contribution of the recollimation shock downstream is less important in the optical band (Fig. \ref{fig:optical emissivity}), which in turn increases the degree of polarization since the macroparticles experience a more coherent magnetic field. For the same reason, the X-ray polarization degree is even higher, since the emission region shrinks further (Fig. \ref{fig:xray emissivity}), resulting almost exclusively in the downstream of the reflection shock. We recover here the main ingredient of the stratified shock scenario.

The emissivity of the reflection shock downstream prevails for higher frequencies for two reasons. First, since the upstream converging flow is almost always perpendicular to the reflection shock, the kinetic energy dissipation is high, with a large amount of internal energy available for the accelerated macroparticles. Second, the magnetic field downstream of the recollimation shock is larger than in the recollimation post-shock region, thus the macroparticles have a larger synchrotron output. 

Simulation C displays the opposite trend, namely the polarization degree decreases with frequency. In this case, the downstream of both the recollimation and reflection shock is dominated by the toroidal component; therefore, the polarization depends exclusively on the asymmetry of the emission region. In fact, since we are observing at a non-negligible angle, relativistic beaming and the $|\vec{B}^\prime \times \hat{\bf n}^\prime|$ term break the jet symmetry. As in the other two simulations, the downstream regions of both the recollimation and reflection shocks contribute to the radio radiative output. Since the radio emission region extends far from the jet axis, the asymmetry becomes more pronounced, leading to a higher degree of polarization. In contrast, the emission region in the optical and X-ray bands is more confined, eventually shrinking to just the downstream region of the reflection shock. Because asymmetry is less significant closer to the jet axis, the overall polarization degree is reduced. 

In \fig{polarization features}, the polarization angle is measured clockwise from the projection of the jet axis onto the sky plane. In Simulation C, the toroidal component of the fluid rest frame magnetic field is dominant by default. As a result, the polarization angle is aligned with the sky projection of the jet axis across all frequencies. In contrast, in Simulation A, the poloidal component of the magnetic field dominates. This causes the polarization angle to be perpendicular to the projection of the jet axis onto the sky plane.

In Simulation B, at high frequencies, the polarization angle is perpendicular to the jet axis projection, while at lower frequencies it is aligned with it. During the transition, the polarization degree temporarily vanishes before increasing again at even lower frequencies. This smooth transition in Simulation B contrasts with the abrupt changes observed in simpler models, such as cylindrical jets (e.g., \citealt{lyutikov05}). Due to the complex structure of the jet flow and the magnetic field in our simulation, the polarization angle can assume intermediate values. The polarization properties of Simulation B can be explained by the structure of the synchrotron emissivity and magnetic fields. The emission region at high frequencies corresponds primarily to the reflection shock downstream, where the poloidal component is significant. At lower frequencies, however, the emission is increasingly influenced by the recollimation shock downstream, where the toroidal component prevails. As a result, the growing contribution of the toroidal component reduces the polarization degree and causes the polarization angle to rotate. Finally, since the toroidal contribution is dominant for even lower frequencies, the polarization angle aligns with the jet axis projection and the polarization degree returns to increase. 

Finally, in appendix \ref{sec:viewing angle}, we report for Simulation A the dependence of the polarization degree as a function of the frequency on the observer viewing angle. The polarization degree varies smoothly across different frequencies as the viewing angle changes. At a fixed frequency, the polarization degree decreases as the viewing angle becomes smaller. At lower viewing angles, the emission region appears more symmetric because the beaming is more uniform. Since the magnetic field is axisymmetric, this results in a reduced polarization degree. However, as the viewing angle changes, the chromaticity between remains almost constant.

\section{Discussion}
\label{sec:discussion}

In this paper, we used the Lagrangian macroparticle approach to probe the nonthermal emission of weakly magnetized blazar jets, specifically to replicate and interpret multifrequency polarimetric observations of HBLs and EHBLs. The Lagrangian approach bridges the gap between fluid-scale structures and kinetic-scale phenomena. While fluid simulations provide bulk velocities, fields, and shock structures at large scales, the macroparticle approach incorporates shock acceleration. 

The first result we found is that shocks from a recollimating jet are characterized by superluminal configurations, limiting the effectiveness of acceleration in purely laminar flows. However, the study of alternative scenarios (explored through PIC simulations) that include turbulence and/or inhomogeneities on kinetic scales demonstrates that acceleration can occur, resulting in nonthermal power law distributions (e.g., \citealt{bresci+23,demidem+23}).

Assuming that acceleration is active, our results for Case B, for which the poloidal and the toroidal components are comparable, prove that the model can easily reproduce the chromaticity of the polarization degree observed in HBLs and EHBLs in quiescent states (e.g., \citealt{liodakis+22,ehlert+23}), even assuming globally ordered fields (i.e., no turbulence at fluid scales). This is in line with the findings of \cite{bolis+24}. Moreover, the model also naturally accounts for the observed overlapping of the optical and X-ray EVPA. 

The model does not reproduce the alignment of the polarization angle with the jet axis observed in some HBL (e.g., \citealt{liodakis+22}). This misalignment can be traced to the intrinsic magnetic field structure in our simulations. The optical and X-ray emission regions almost coincide with the reflection shock downstream. Specifically, the emission is higher around the $z$ axis, where the toroidal magnetic field component is zero by definition, leaving the poloidal component dominant at these frequencies. As a result, the polarization angle is oriented perpendicular to the projected jet axis. A jet with a dominant toroidal component (case C) displays the EVPA aligned with the projected jet axis, but the properties of the degree of polarization (decreasing with increasing frequencies) are incompatible with the observations. It is worth noting here that the alignment is evident for Mrk501 \citep{liodakis+22}, while it appears less definitive for Mrk421 \citep{digesu+22} and 1ES 0229+200 \citep{ehlert+23}. At any rate, the mismatch could be explained by the jet bending between the inner X-ray and optical emission region and the outer extended region observed by radio telescopes \citep{digesu+22}. Moreover, jets can change direction in time \citep{kostrichkin+24}, thus requiring simultaneous radio imaging for an appropriate estimate of the alignment of the polarization angles with the jet projection.

Our results could depend on the magnetic field geometry since different profiles (e.g., \citealt{hu+25}) could lead to subluminal shocks or different polarization features. We plan to explore different configurations in a future paper. 

Our current post-processing code has so far been applied only to axisymmetric RMHD simulations without accounting for the system time evolution after performing 3D interpolation. As previously discussed, fluid-level turbulence can arise downstream of the reflection shock. While particle acceleration at shocks should not be affected, since in fully time-evolved 3D simulations the reflection shock remains largely intact (\citealt{costa23}, Boula et al. in prep.), the turbulent magnetic field of the reflection shock downstream could affect the emission. This could help to relax the fine-tuning of the pitch angle needed to reproduce observations. While Simulation B closely reproduces the IXPE and multiwavelength data, in Simulation A the polarization degree is too large and not enough chromatic. However, a more disordered magnetic field would reduce the overall polarization degree. Moreover, turbulence should have a greater impact on electrons that travel longer distances (namely emitting at lower frequencies), accentuating the chromaticity. Finally, although our present code only explains the polarization of blazars in a quiescent state, incorporating the time evolution and subsequent turbulence-driven symmetry breaking could help explain additional phenomena, such as the observed X-ray polarization angle swing in Mrk 421. Finally, turbulence may also contribute to particle acceleration. The current spectral update implementation is already designed to incorporate stochastic acceleration by turbulence (see also \citealt{kundu+21}). 

Another source of turbulence could be intrinsic to the jet flow. At fluid scales, our RMHD simulations display a laminar flow. However, the pre-shock fluid could be inhomogeneous and/or turbulent even on fluid scales. The non-laminarity of the fluid could result in a more turbulent downstream, leading also to magnetic field amplification \citep{giacalone&jokipii07, mizuno+14}. As a result, the emission from the macroparticles could change, especially the predicted polarization degree. Moreover, strong turbulence could also further accelerate particles, as outlined in the previous paragraph. 

It is also important to note that the framework we have described is well-suited for HBLs and EHBLs. For other classes of blazars, such as FSRQs, LBLs, and IBLs, the situation is more complex. These sources tend to have higher luminosities and lower characteristic peak frequencies, which implies that the conditions in the emitting regions differ significantly. They require larger magnetic fields and a higher nonthermal electron number density. Under such conditions, the maximum Lorentz factor for which the cooling length is higher than the decay length of the microturbulence, see \eq{cooling vs decay}, could be much lower, making our approximation inconsistent in the bands of interest. Moreover, the ratio of the high and low energy peaks is higher than for HBLs and EHBLs, thus the contribution of the inverse Compton cooling cannot be neglected for these sources. We plan to implement inverse Compton cooling and emission in future work (for a related example of external Compton processes on the cosmic microwave background, see \citealt{vaidya+18}).

\begin{acknowledgements}
We thank E. Sobacchi for fruitful discussions. We acknowledge financial support from an INAF Theory Grant 2022 (PI F.~Tavecchio). This work has been funded by the European Union-Next Generation EU, PRIN 2022 RFF M4C21.1 (2022C9TNNX). This work has been funded by ASI under contract 2024-11-HH.0. We acknowledge support by CINECA, through ISCRA and Accordo Quadro INAF-CINECA, and by PLEIADI, INAF – USC VIII, for the availability of HPC resources. Plots have been made with Python. We have used the following libraries: Numpy \citep{numpy}, Matplotlib \citep{matplotlib}, Scipy \citep{scipy}, and PyPluto \citep{pypluto25}.
\end{acknowledgements}

\bibliographystyle{aa}
\bibliography{bibliography}

\appendix

\section{Derivation of the magnetic field profile \label{sec:B_details}}
In this work, we simulated a helical magnetic field in a conical jet. In such a case, the field is made both of a toroidal field, perpendicular to the streamlines, and of a poloidal component, directed along the spherical radius $r$; no component is null in cylindrical coordinates: $\mathbf{B} = \left(B_R,B_z,B_\phi\right)$. We remind that a factor $\sqrt{4\pi}$ is reabsorbed in the definition of the magnetic field and that we work with $c=1$. To find a simple way to prescribe analytically the magnetic field, it is easier to work in spherical coordinates $(r,\theta,\phi)$, because $\mathbf{v} = \left(v_r,0,0\right).$ We can assume that $B_\theta=0$, so that the field is $\mathbf{B} = \left(B_r,0,B_\phi\right)$; the poloidal and toroidal components are simply $B_{pol}=B_r$ and $B_{tor}=B_\phi$, and due to axisymmetry none of them depend on $\phi.$ These quantities are defined in the laboratory-frame; the proper frame magnetic field is $\mathbf{B}'=(B_r,0,B_\phi/\Gamma)$. While the proper-frame electric field is null in ideal magneto-hydrodynamics, the laboratory frame electric field is given by 
\begin{equation}
    \mathbf{E} = -\mathbf{v\times B} = v_r B_\phi \hat{e}_\theta.
    \label{eq:electric}
\end{equation}

In ideal MHD, the magnetic flux is conserved. Since the jet's initial (and boundary) geometry is conical, its section is $\Sigma \propto r^2$ and this determines a condition on the poloidal field
\begin{align}
    \int_{r^2} \mathbf{B}\cdot d\mathbf{S}  = \text{conserved}\rightarrow B_{r} = \frac{g(\theta)}{r^2};
\end{align}
on the other hand, the conservation of current implies
\begin{align}
& I = \oint_{\partial \Sigma} \mathbf{B}\cdot d\mathbf{l} - \frac{d}{dt}\int_\Sigma \mathbf{E}\cdot d\mathbf{S} \propto \oint_{r} \mathbf{B}\cdot d\mathbf{l} - k\frac{d}{dt}\int_{r^2} \mathbf{E}\cdot d\mathbf{S},\\
&\rightarrow B_\phi=\frac{h(\theta)}{r},
\end{align}
where $k$ is a constant and the displacement current is zero because $E_r = 0$.
The dependence on the spherical radius is different for the two magnetic field components, so it is not possible to define a magnetic field in an equilibrium configuration in the whole conical jet. Since the initial configuration is merely a starting point, and the only region where the initial condition is maintained throughout the simulation is the nozzle, we limit to setting the equilibrium across the jet (in the $\theta$ direction) at its injection in $(r = 1,\theta)$:
\begin{equation}
    \left[\left[\mathbf{(\nabla\cdot E)E}+\mathbf{(\nabla\times B)\times B}\right]^\theta-\frac{\partial p}{\partial \theta}\right ](r=1,\theta)=0.
    \label{eq:f0}
\end{equation}
Making explicit the pressure dependency on the spherical radius, we have $p=\left(\tilde{p}(\theta)+\tilde{p}_0\right) r^{-2\gamma}$, where $\tilde{p}_0$ is a constant, and $\gamma_\text{ad}=5/3$ is the adiabatic index. Then, using also eq. \ref{eq:electric}, we can solve the previous equation,
\begin{gather}
    \frac{\partial}{\partial \theta}\left(\frac{h^2(\theta)}{\Gamma^2}\right)+ 2 \cot{\theta}\, \left(\frac{h^2(\theta)}{\Gamma^2}\right) = -\frac{\partial g^2(\theta)}{\partial \theta}-\frac{\partial \tilde{p}(\theta)}{\partial \theta} ,\\
     \frac{\partial}{\partial \theta}\left(\frac{h^2(\theta)}{\Gamma^2}\right)+ 2 \cot{\theta}\, \left(\frac{h^2(\theta)}{\Gamma^2}\right) = -\frac{\partial l^2(\theta)}{\partial \theta} ,\label{eq:f0_final}
\end{gather}
where we defined for simplicity $l^2(\theta)= g^2(\theta)+\tilde{p}(\theta)$.  

\begin{figure}[htbp]
    \centering
    \includegraphics[width=\columnwidth]{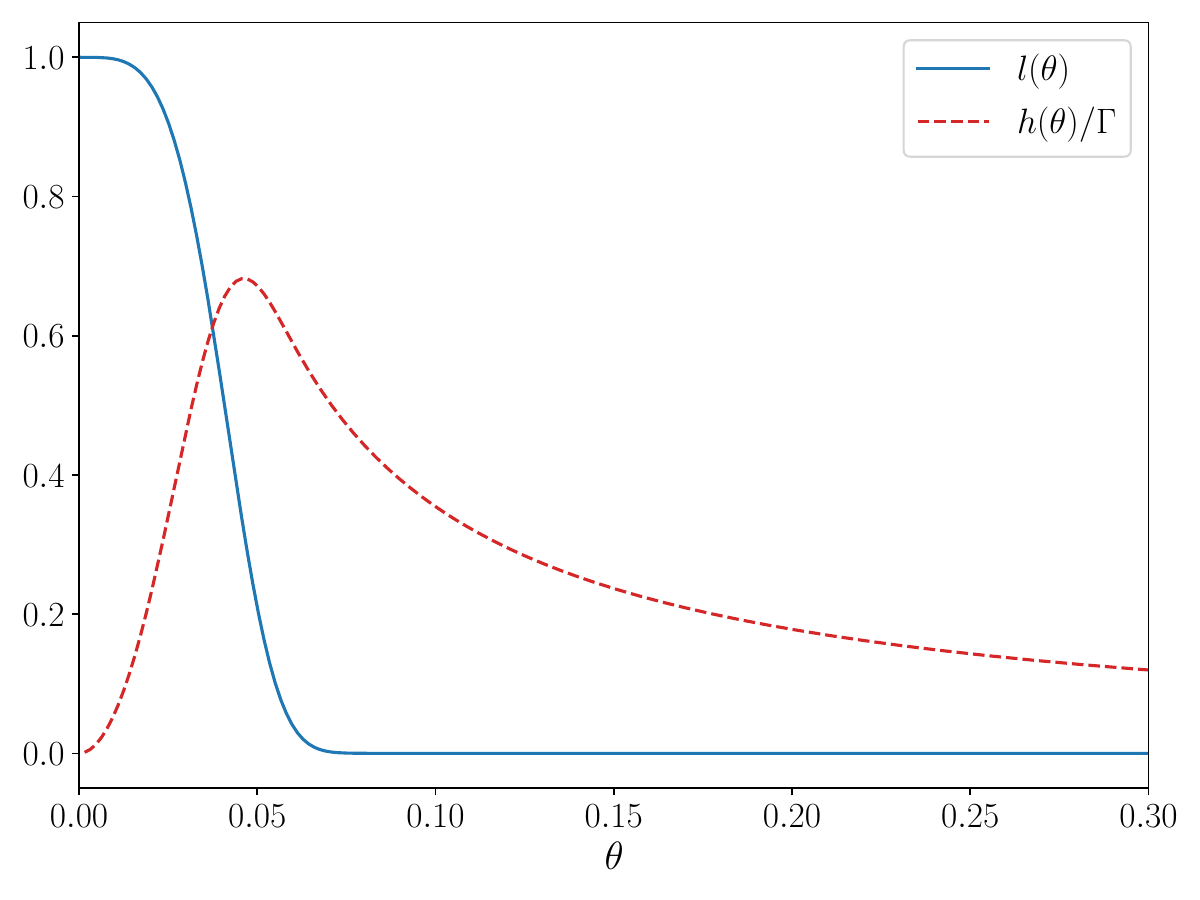}
    \caption{Profiles of $l(\theta)$ and $h(\theta)/\Gamma$.}
    \label{fig:B_profiles}
\end{figure}

If we adopt the following Gaussian profile $l^2(\theta) = e^{-2\mathcal{X}^2}$, where $\mathcal{X} = \left(\cos\theta-1\right)/\psi_\chi$ and $\psi_\chi$ is an arbitrary constant, equation \ref{eq:f0_final} is solved by
\begin{align}
    h^2(\theta) = \Gamma^2 \left[-\text{e}^{-2\mathcal{X}^2} -\frac{\psi_\chi}{2 \sin^2\theta}\left(\psi_\chi-\psi_\chi \text{e}^{-2\mathcal{X}^2}+2\sqrt{\pi}\,\erf\left(\sqrt{2}\mathcal{X}\right)\right)\right].
\end{align}
The functions $l(\theta)$ and $h(\theta)$ yield appropriate profiles for the poloidal and toroidal field components if $\psi_\chi = 10^{-2}\theta_j$, as shown in Fig. \ref{fig:B_profiles} for $\theta_j=0.1$: $\max[l(\theta)]=l(0)=1$, $h(0)=0,$ $\max[h(\theta)]\simeq h(0.05)\simeq0.7\, \Gamma$, and both functions decay to $0$ for $\theta>\theta_j=0.1$. 

We define the lab-frame toroidal field as 
\begin{equation}
    B_\phi = \frac{B_0\Gamma}{r}\sqrt{-\text{e}^{-2\mathcal{X}^2} -\frac{\psi_\chi}{2 \sin^2\theta}\left(\psi_\chi-\psi_\chi \text{e}^{-2\mathcal{X}^2}+\sqrt{2\pi}\,\erf\left(\sqrt{2}\mathcal{X}\right)\right)},
    \label{eq:bphi}
\end{equation}
and the poloidal field as
\begin{equation}
    B_r = \alpha B_0 \frac{e^{-\mathcal{X}^2}}{r^2},
    \label{eq:br}
\end{equation}
with the parameter $\alpha$ that determines the pitch. The pressure profile is chosen to satisfy equation \ref{eq:f0_final},
\begin{equation}
    \tilde{p}(\theta) = (1-\alpha^2) B^2_0 e^{-2\mathcal{X}^2} \rightarrow p(r,\theta) = (1-\alpha^2)\frac{B^2_0}{2}  \frac{\text{e}^{-2\mathcal{X}^2}}{r^{2\gamma_\text{ad}}}+\frac{p_{j,0}}{r^{2\gamma_\text{ad}}},
\end{equation}
where the constant $p_{j,0}$ is chosen as floor jet pressure value at injection.

In the previous calculations, we found a magnetic field configuration that is in equilibrium in the $\theta$ direction at $(r=1,\theta)$. Instead, simulations are performed in a cylindrical domain, and the jet injection boundary corresponds to the region $(R<\theta_j,z=1)$, where $R=r\sin\theta$ and $z=r\cos\theta$. For small values of $\theta$, within the jet, where the profile is relevant, we can approximate the two equilibrium states, since $R(r=1,\theta)\sim\theta$ and $z(r=1,\theta)\sim1$, so we inject the cylindrical field 
\begin{equation}
    \mathbf{B}(R,z)=\left(B_r(R,z) \frac{R}{\sqrt{R^2+z^2}},B_\phi(R,z),B_r(R,z)\frac{z}{\sqrt{R^2+z^2}}\right).
\end{equation}

\section{Numerical scheme for spectral update \label{sec:spectral update scheme}}

In our code, the numerical scheme for the spectral update step is a second-order implicit-explicit Runge-Kutta scheme, specifically the strong stability preserving algorithm (2,2,2), see \cite{pareschi&russo05}. The choice of this scheme permits the future introduction of a diffusion term accounting for turbulence acceleration, as in \cite{kundu+21}. Since the advection term is solely present, the scheme reduces to 
\begin{equation}
    \left \{
    \begin{aligned}        
        & \chi_{i,*} = \chi_{i,n} + \Delta t_n \mathcal{A}_{i,n},  \\
        & \chi_{i,n+1} = \chi_{i,n} + \frac{\Delta t_n}{2} \left [ \mathcal{A}_{i,n} + \mathcal{A}_{i,*} \right].
    \end{aligned}
    \right .
\end{equation}
The subscript $i$ indicates the $i$-th point in the logarithmically equally spaced energy grid, while $\mathcal{A}$ is the advection term, discretized as 
\begin{equation}
\mathcal{A}_{i,n} = -\xi'_i \frac{\mathcal{F}^{\text{adv}}_{i+\frac{1}{2},n} - \mathcal{F}^{\text{adv}}_{i-\frac{1}{2},n}}{\Delta \xi},
\end{equation}
where $\xi_i = \log(E_i/E_\mathrm{min})/\log(E_\mathrm{max}/E_\mathrm{min})$ and $\xi'_i = [E_i \log(E_\mathrm{max}/E_\mathrm{min})]^{-1}$ is the corresponding Jacobian. $\mathcal{F}^{\text{adv}}$ indicates the advection flux, discretized as
\begin{equation}
    \mathcal{F}_{i+\frac{1}{2},n}^{\text{adv}} =
    \begin{cases}
    H(\gamma_{i+\frac{1}{2}}) \chi_{i+\frac{1}{2},n}^L & \text{if } H(\gamma_{i+\frac{1}{2}}) > 0, \\
    H(\gamma_{i+\frac{1}{2}}) \chi_{i+\frac{1}{2},n}^R & \text{if } H(\gamma_{i+\frac{1}{2}}) < 0,
    \end{cases} 
\end{equation}
where $H$ is the sum of the two terms in the round brackets of \eq{simplified transport equation 3}. Finally, the left $\chi^L$ and right $\chi^R$ are defined as
\begin{gather}
    \chi_{i+\frac{1}{2}}^L = \chi_i + \frac{\delta \chi_i}{2}, \\
    \chi_{i+\frac{1}{2}}^R = \chi_{i+1} - \frac{\delta \chi_{i+1}}{2}, \\
    \text{with } \delta \chi_i =
    \begin{cases}
    \frac{2 \Delta \chi_{i+\frac{1}{2}} \Delta \chi_{i-\frac{1}{2}}}{\Delta \chi_{i+\frac{1}{2}} + \Delta \chi_{i-\frac{1}{2}}} & \text{if } \Delta \chi_{i+\frac{1}{2}} \Delta \chi_{i-\frac{1}{2}} > 0 \\
    0, & \text{otherwise},
    \end{cases}
\end{gather}
where $\Delta \chi_{i \pm \frac{1}{2}} = \pm \left( \chi_{i \pm 1} - \chi_i \right)$. Note that for this scheme two ghost cells are needed on both sides of the energy grid. To conserve particle number, a zero flux condition must be imposed at both borders. For our algorithm, this translates into the following boundary conditions
\begin{gather}
    \begin{aligned}
    \chi_i^{\text{adv}} &=
    \begin{cases}
    -\chi_{2i_b - i - 1}^{\text{adv}}, & \text{for } i < i_b, \\
    -\chi_{2i_e - i + 1}^{\text{adv}}, & \text{for } i > i_e,
    \end{cases} 
    \end{aligned} \\
    \text{with } \mathcal{F}_{i_b - \frac{1}{2}}^{\text{adv}} = \mathcal{F}_{i_e + \frac{1}{2}}^{\text{adv}} = 0,
\end{gather}
where $i_b$ and $i_e$ are the indices of the first and last points of the energy grid, respectively. This scheme is second-order accurate, but the Courant–Friedrichs–Lewy condition ($C = \mathcal{A} \Delta t/\Delta E <1$, with $C$ the Courant number) must be satisfied since the advection term is treated explicitly. 

\section{From laboratory to upstream and wall frame: detailed implementation \label{sec:frame transformations}}
To boost to the wall and upstream rest frames, we first compute the shock normal in the laboratory frame $\hat{\bf n}_\mathrm{sh}$ by means of the coplanarity theorem (it can be proved directly from the RMHD jump conditions), which states that the upstream and downstream magnetic field along with the velocity jump lie in the same plane as the shock normal \citep{landau&lifshitz60},
\begin{equation}
    \hat{\bf n}_\mathrm{sh} =
    \left \{
    \begin{aligned} 
    &\pm \frac{\vec{\beta}_2 - \vec{\beta}_1}{|\vec{\beta}_2 - \vec{\beta}_1|} \quad \text{parallel shocks} , \\
    &\pm \frac{(\vec{B}_1 \times \Delta \vec{\beta}) \times \Delta \vec{B}}{|(\vec{B}_1 \times \Delta \vec{\beta}) \times \Delta \vec{B}|} \quad \text{otherwise},
    \end{aligned}
    \right . 
\end{equation}
where the subscripts $1$ and $2$ indicate, respectively, the upstream and downstream values. Note that a different routine is implemented for parallel shocks since $\Delta \vec{B} = 0$. Then, by enforcing number flux conservation across the shock surface, we calculate the shock speed in the laboratory frame, whose direction is along the shock normal:
\begin{equation}
    \beta_\mathrm{sh} = \frac{\rho_2 \gamma_2 \vec{\beta}_2 \cdot \hat{\bf n}_\mathrm{sh} - \rho_1 \gamma_1 \vec{\beta}_1 \cdot \hat{\bf n}_\mathrm{sh}}{\rho_2 \gamma_2 - \rho_1 \gamma_1}.
\end{equation}
Given the shock normal and speed, the fluid quantities can be transformed to the normal incident frame (NIF), where the shock is at rest and the upstream velocity is directed along the shock normal. This transformation requires a two-step procedure, in analogy with the process outlined in \cite{kirk&heavens89} and \cite{summerlin&baring12} for the de Hoffmann-Teller frame. First, we perform a Lorentz boost equal to the shock speed and along the shock normal. Then, the second boost is along the transverse component of the upstream velocity. The two-step procedure has an important advantage: the shock normal does not vary across the three different frames, namely $\hat{\bf n}_\mathrm{sh} = \hat{\bf n}_\mathrm{sh,int} = \hat{\bf n}_\mathrm{sh, NIF}$, where the subscript ``int'' indicates the intermediate frame between the laboratory and NIF. The first boost is along the shock normal, which therefore remains unchanged, while the second boost, though perpendicular to the normal, also leaves the normal unchanged, since the shock is at rest. Defining $\Lambda[\vec{u}]$ as a generic Lorentz boost along the velocity $\vec{u}$, a four-vector $\mathbf{x}$ in the laboratory frame can be transformed to the NIF as follows
\begin{equation}
    \mathbf{x}_\mathrm{NIF} = \Lambda [ \vec{v}_\mathrm{1, int} - (\vec{v}_\mathrm{1, int} \cdot \hat{\bf n}_\mathrm{sh}) \hat{\bf n}_\mathrm{sh}]\, \Lambda[v_\mathrm{sh} \hat{\bf n}_\mathrm{sh}]\, \mathbf{x}, 
    \label{eq:NIF transformation}
\end{equation} 

We can transform to the upstream rest frame with a further step. We perform a boost along the NIF upstream velocity, which is parallel to the shock normal:
\begin{equation}
    \mathbf{x}_\mathrm{up} = \Lambda[\vec{v}_\mathrm{1,NIF}]\, \mathbf{x}_\mathrm{NIF}.
\end{equation}
The boost is again along the shock normal, which, therefore, remains unchanged. Hence, we can substitute $\hat{\bf n}_\mathrm{sh, up}$ with $\hat{\bf n}_\mathrm{sh}$ in the superluminal condition.

Similarly, it is possible to transform from the NIF to the wall frame in one step, with a single boost along the normal component of the downstream velocity: 
\begin{equation}
    \mathbf{x}_\mathrm{wall} = \Lambda[(\vec{v}_\mathrm{2,NIF} \cdot \hat{\bf n}_\mathrm{sh})\hat{\bf n}_\mathrm{sh}]\, \mathbf{x}_\mathrm{NIF}.
\end{equation}
As before, the shock normal does not vary, namely $\hat{\bf n}_\mathrm{sh, wall} = \hat{\bf n}_\mathrm{sh}$. Moreover, the upstream velocity remains along the shock normal. Note that, since the magnetization is low, it is possible to prove that the transverse component of the downstream velocity is nonrelativistic; therefore, no extra boost is needed to move from the wall frame to the downstream rest frame. 

Transforming from the laboratory to the NIF, upstream, and downstream rest frame may seem unnecessarily complex. A single Lorentz boost appears to be the simplest approach for the upstream and downstream rest frames. However, this approach results in a rotation of the shock normal \citep{kirk&heavens89, ballard&heavens91}. While this difference in the shock normal behavior may seem like a discrepancy, it is actually a direct consequence of the well-known Thomas-Wigner rotation: the composition of two non-collinear boosts results in a Lorentz transformation that is not a pure boost but is the composition of a boost and a rotation. To avoid complications arising from this effect, we have opted for the two-step procedure.

\onecolumn

\section{Subluminal versus superluminal scenario in Simulation B and C \label{sec:versus B and C}}

\begin{figure*}[htbp]
\centering
\includegraphics[width=.97\textwidth]{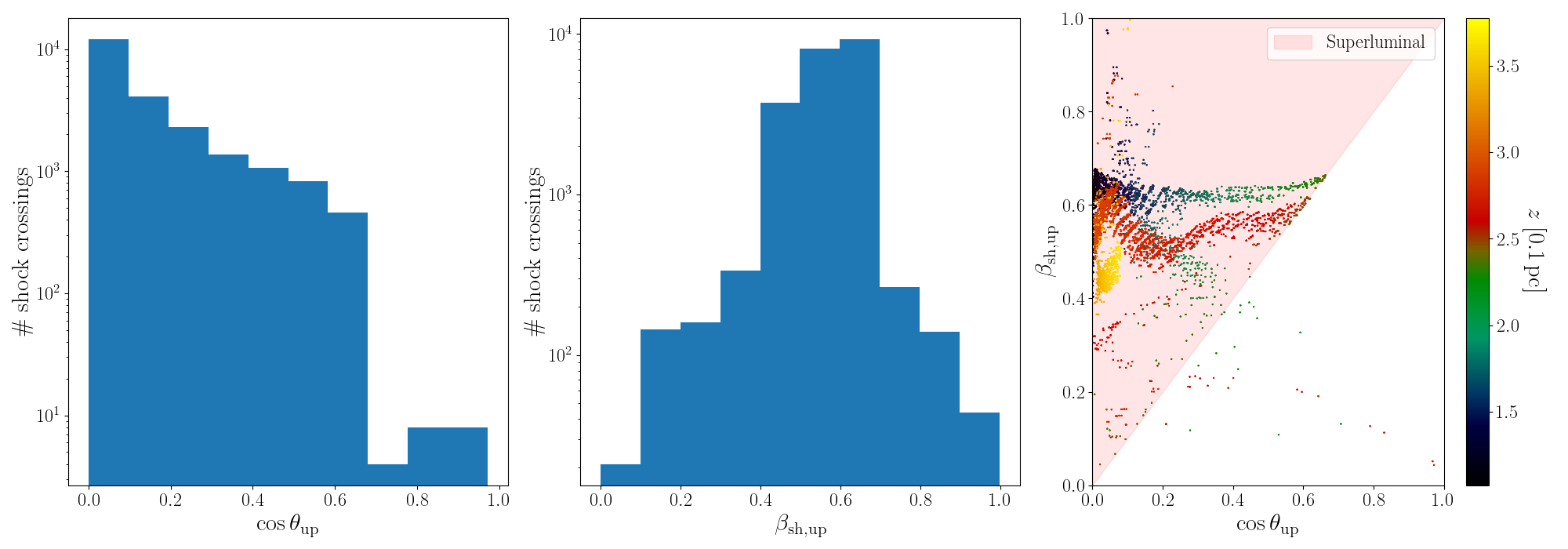}
\caption{Same as \fig{super vs sub A} but for Simularion B. See the caption of \fig{super vs sub A} for details.}
\label{fig:super vs sub B}
\end{figure*}

\begin{figure*}[htbp]
\centering
\includegraphics[width=.97\textwidth]{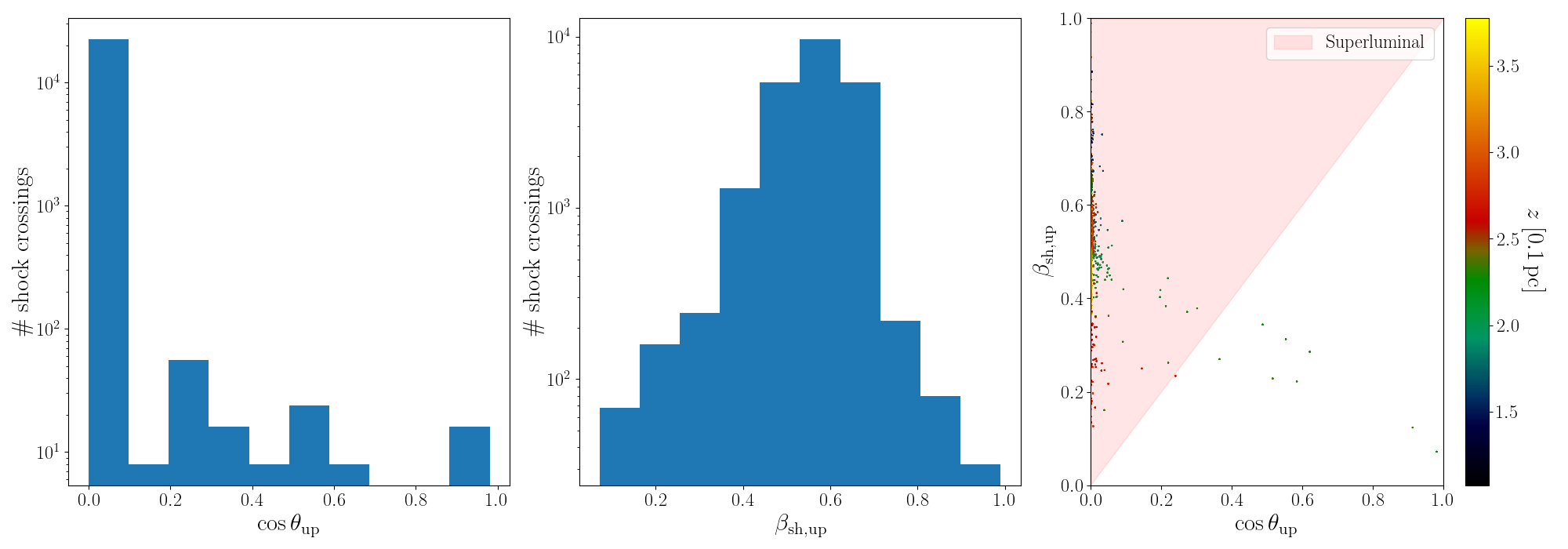}
\caption{Same as \fig{super vs sub A} but for Simularion C. See the caption of \fig{super vs sub A} for details.}
\label{fig:super vs sub C}
\end{figure*}

\FloatBarrier
\twocolumn

\section{Simulation A, B and C fluxes \label{sec:sim fluxes}}

\begin{figure}[htbp]
\centering
\includegraphics[width=\columnwidth]{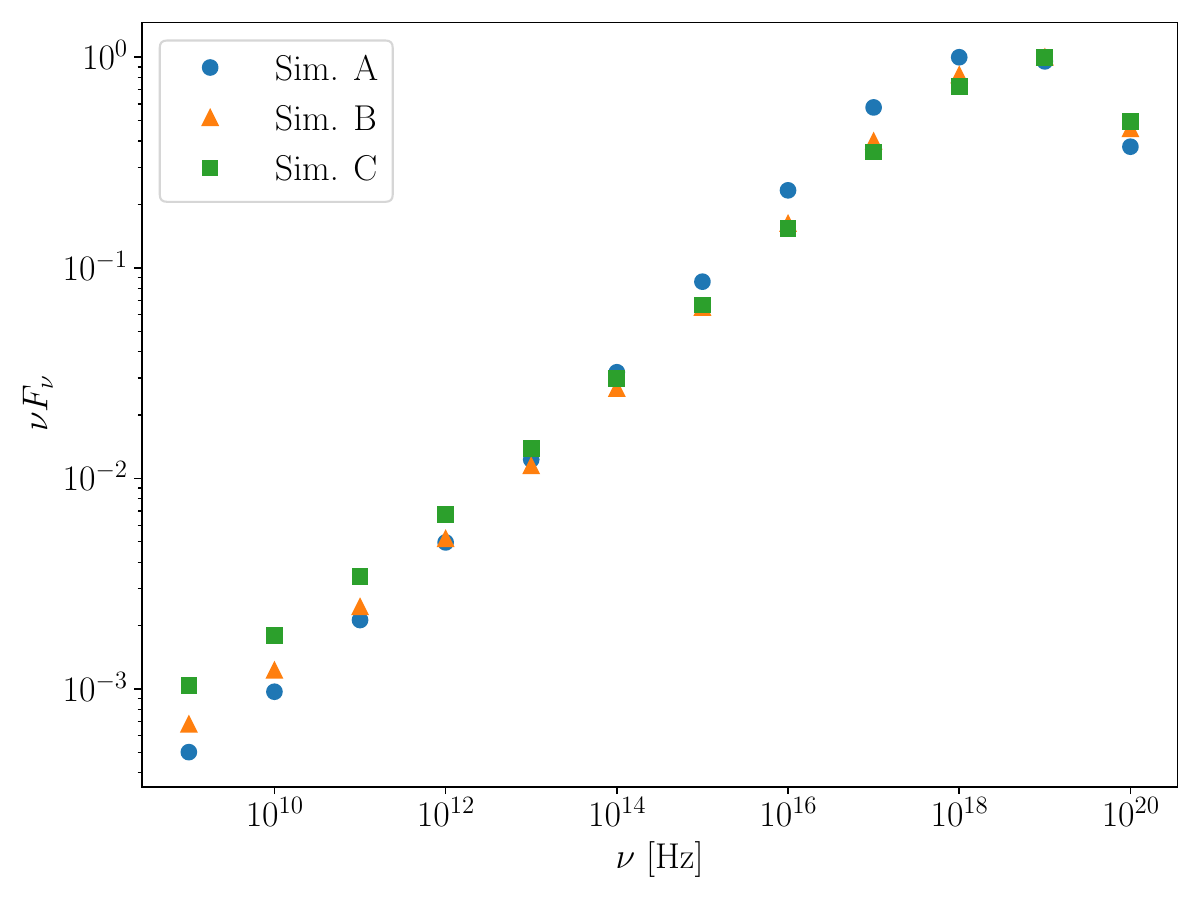}
\caption{Normalized observer frame synchrotron flux of Simulation A (blue circles), B (orange triangles), and C (green squares) as a function of frequency.}
\end{figure}

\section{Simulation A emissivities \label{sec:sim A emissivities}}

\begin{figure}[htbp]
\centering
\includegraphics[width=\columnwidth]{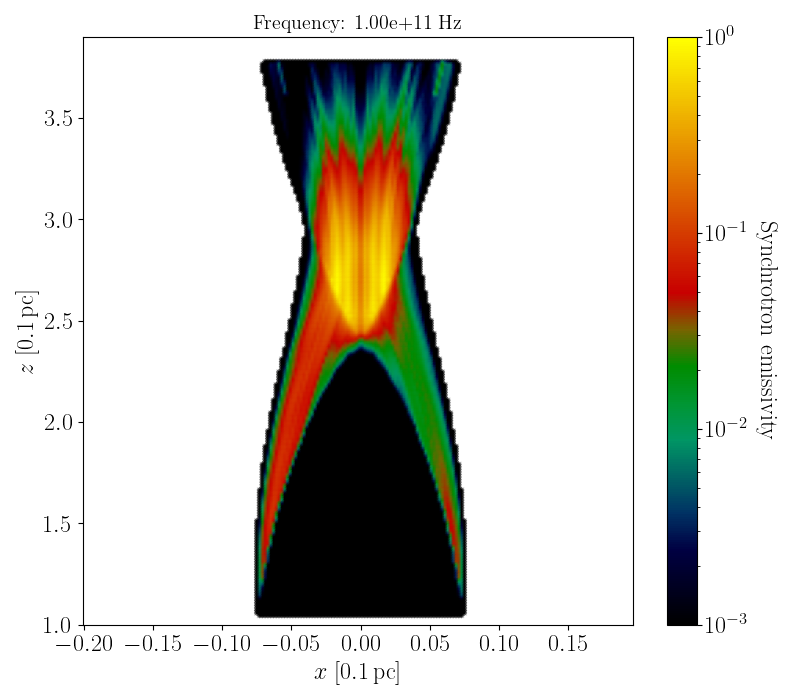}
\caption{Normalized observer frame synchrotron emissivity in the radio band of a 10-code units slab centered on $xz$ plane (Simulation A).}
\label{fig:radio emissivity A}
\end{figure}

\begin{figure}[htbp]
\centering
\includegraphics[width=\columnwidth]{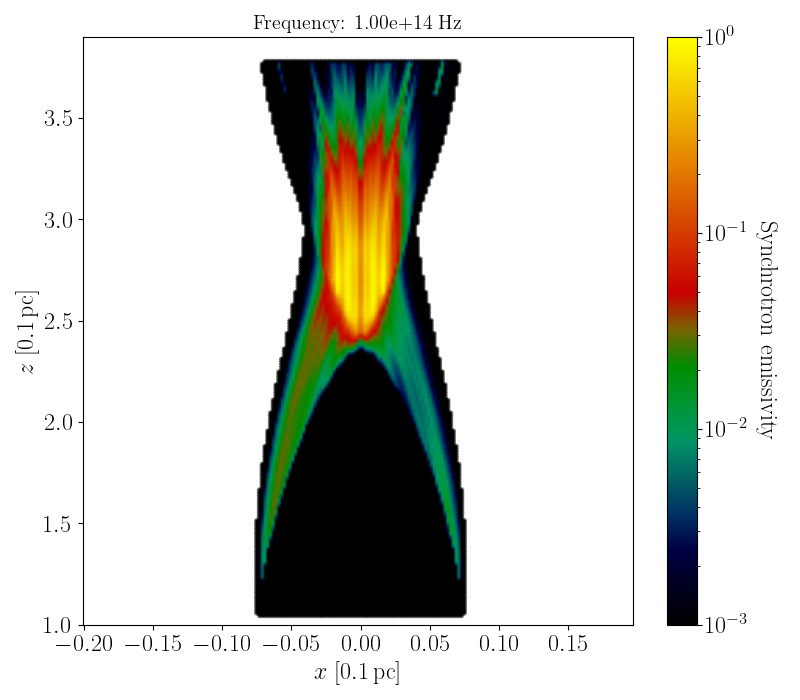}
\caption{Same as Fig. \ref{fig:radio emissivity A} but in the optical band.}
\end{figure}

\begin{figure}[htbp]
\centering
\includegraphics[width=\columnwidth]{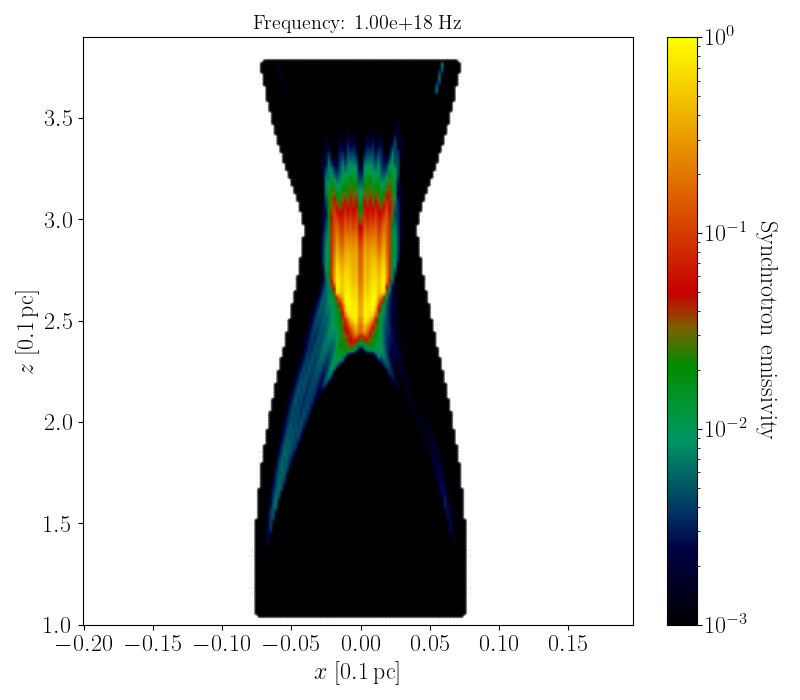}
\caption{Same as Fig. \ref{fig:radio emissivity A} but in the X-ray band.}
\end{figure}

\FloatBarrier
\clearpage

\section{Simulation C emissivities \label{sec:sim C emissivities}}

\begin{figure}[htbp]
\centering
\includegraphics[width=\columnwidth]{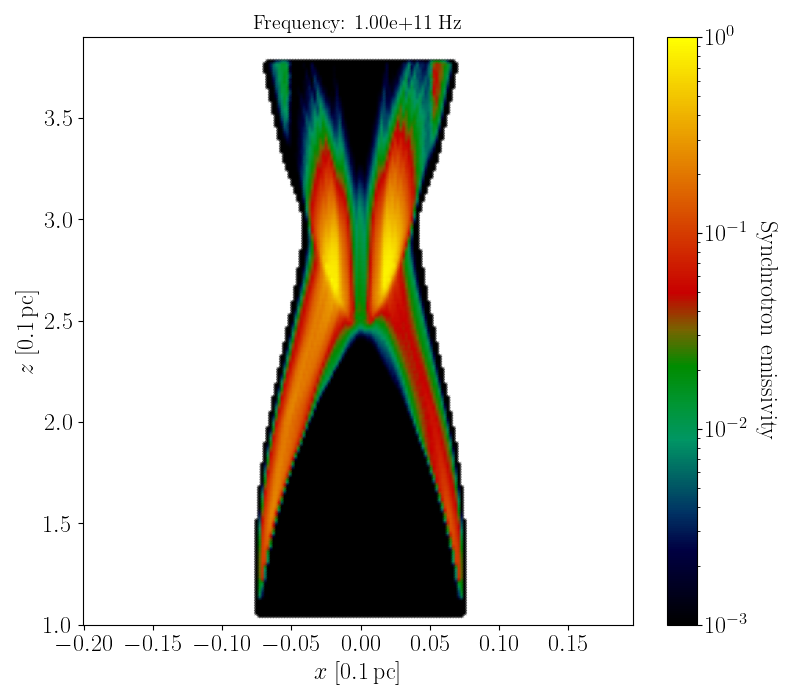}
\caption{Normalized observer frame synchrotron emissivity in the radio band of a 10-code units slab centered on $xz$ plane (Simulation C).}
\label{fig:radio emissivity C}
\end{figure}

\begin{figure}[htbp]
\centering
\includegraphics[width=\columnwidth]{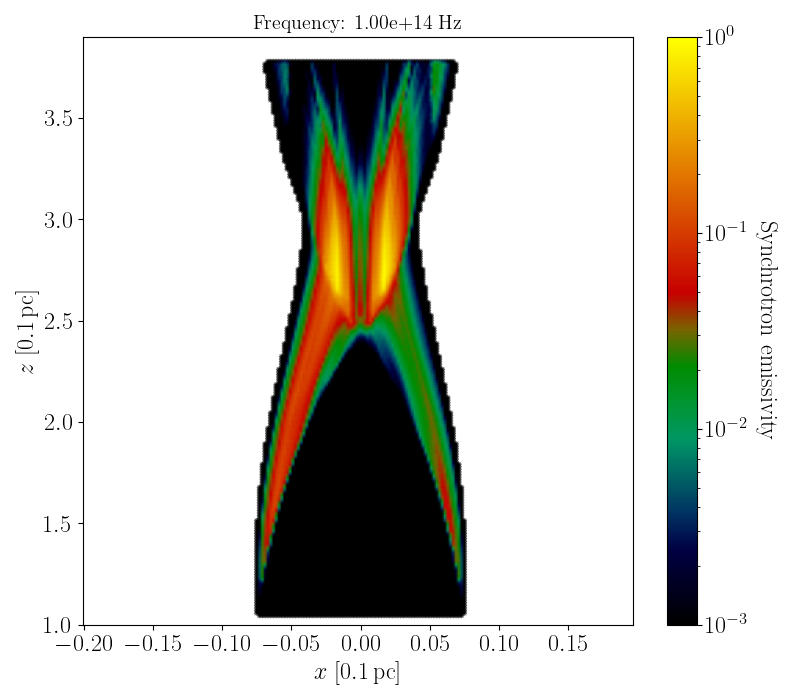}
\caption{Same as Fig. \ref{fig:radio emissivity C} but in the optical band.}
\end{figure}

\begin{figure}[htbp]
\centering
\includegraphics[width=\columnwidth]{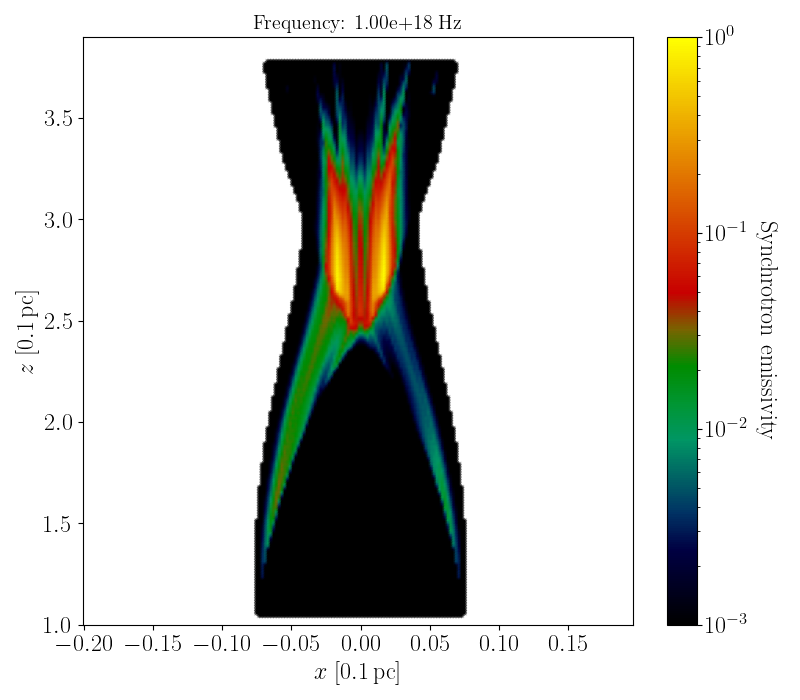}
\caption{Same as Fig. \ref{fig:radio emissivity C} but in the X-ray band.}
\end{figure}

\FloatBarrier

\section{Effect of the viewing angle on the polarization degree \label{sec:viewing angle}}

\begin{figure}[htbp]
\centering
\includegraphics[width=\columnwidth]{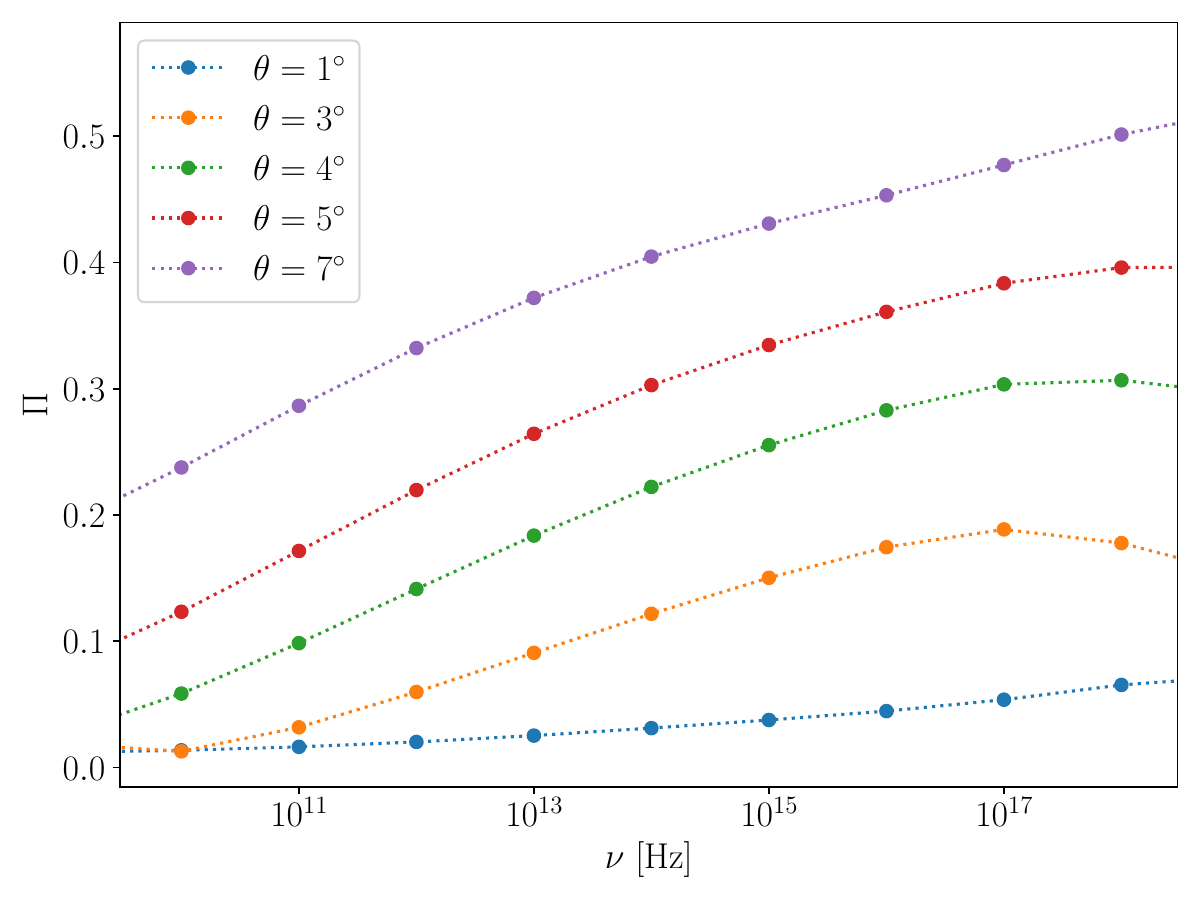}
\caption{Polarization degree of Simulation A as a function of frequency for different viewing angles.}
\end{figure}

\FloatBarrier

\end{document}